\documentclass[prb,twocolumn,superscriptaddress,showpacs,amsmath,amssymb]{revtex4-1}
\usepackage{graphicx}
\usepackage{dcolumn}
\usepackage{bm}
\usepackage{epstopdf}
\usepackage{upgreek}
\usepackage{color}

\usepackage{amsmath}
\usepackage{amssymb}
\usepackage{amsthm}
\usepackage{pinlabel}
\usepackage{natbib}

\newcommand{\ph}{^{\phantom{\dagger}}}
\usepackage{braket}

\begin{document}

\title{Evolution of multi-gap superconductivity in the atomically thin limit: Strain-enhanced three-gap superconductivity in monolayer MgB$_2$} 

\author{J. Bekaert}
\email{jonas.bekaert@uantwerpen.be}
\affiliation{%
 Department of Physics, University of Antwerp,
 Groenenborgerlaan 171, B-2020 Antwerp, Belgium
}%
\author{A. Aperis}
 \affiliation{%
Department of Physics and Astronomy, Uppsala University,
Box 516, SE-751 20 Uppsala, Sweden
}
 \author{B. Partoens}
 \affiliation{%
 Department of Physics, University of Antwerp,
 Groenenborgerlaan 171, B-2020 Antwerp, Belgium
}%
\author{P. M. Oppeneer}
 \affiliation{%
Department of Physics and Astronomy, Uppsala University,
Box 516, SE-751 20 Uppsala, Sweden
}
 \author{M. V. Milo\v{s}evi\'{c}}
\email{milorad.milosevic@uantwerpen.be}
\affiliation{%
 Department of Physics, University of Antwerp,
 Groenenborgerlaan 171, B-2020 Antwerp, Belgium
}

\date{\today}

\begin{abstract}
\noindent Starting from first principles, we show the formation and evolution of superconducting gaps in MgB$_2$ at its ultrathin limit. Atomically thin MgB$_2$ is distinctly different from bulk MgB$_2$ in that surface states become comparable in electronic density to the bulk-like $\sigma$- and $\pi$-bands. Combining the \textit{ab initio} electron-phonon coupling with the anisotropic Eliashberg equations, we show that monolayer MgB$_2$ develops three distinct superconducting gaps, on completely separate parts of the Fermi surface due to the emergent surface contribution. These gaps hybridize nontrivially with every extra monolayer added to the film, owing to the opening of additional coupling channels. Furthermore, we reveal that the three-gap superconductivity in monolayer MgB$_2$ is robust over the entire temperature range that stretches up to a considerably high critical temperature of 20 K. The latter can be boosted to $>$50 K under biaxial tensile strain of $\sim$ 4\%, which is an enhancement stronger than in any other graphene-related superconductor known to date. 
\end{abstract}

\pacs{74.20.Fg,74.20.Pq,74.25.Kc,74.70.Ad,74.78.-w}
\maketitle

\begin{figure*}[ht]
\centering
\includegraphics[width=\linewidth]{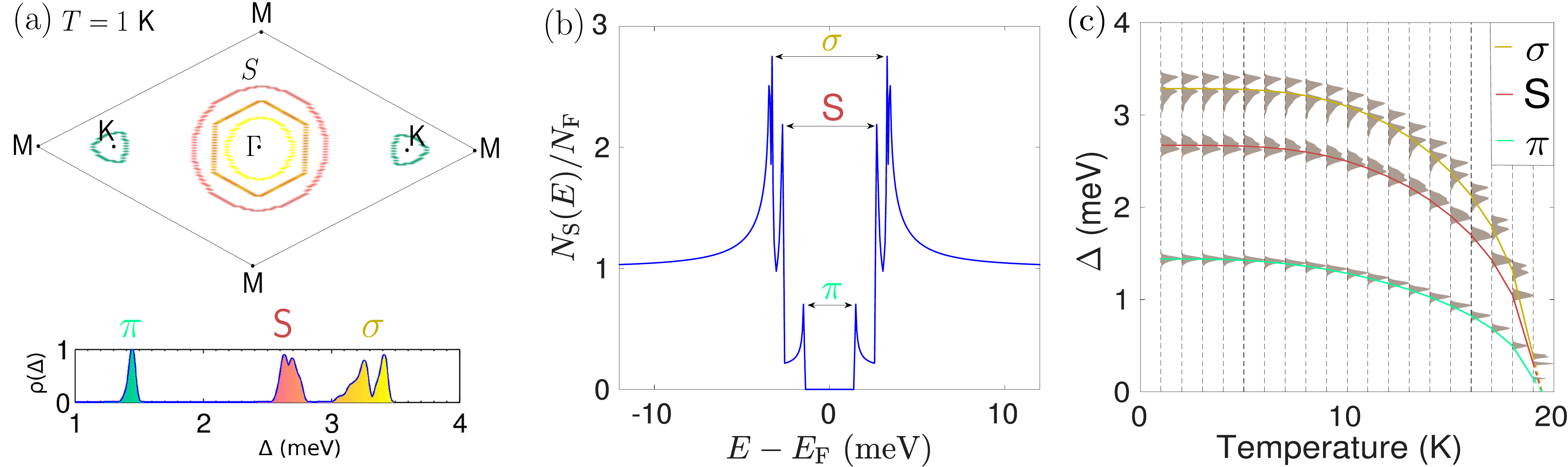}
\caption{(Color online) The superconducting spectrum of 1-ML MgB$_2$, calculated by anisotropic Eliashberg theory with \textit{ab initio} input. (a) The distribution of the three superconducting gaps $\Delta(\textbf{k}_{\mathrm{F}},T)$ on the Fermi surface: $\pi$, \textit{S} (for surface) and $\sigma$, at $T=1$ K. (b) The density of states in the superconducting state at $T=1$ K, showing three distinct peaks corresponding to the three gaps. (c) The evolution of the gap spectrum with temperature, including the gap averages. The calculation shows that 1-ML MgB$_2$ has $T_{\mathrm{c}}\cong 20$ K.}
\label{fig:fig1}
\end{figure*}

\section{Introduction}

A multi-gap superconductor is characterized by separate superconducting gaps opening on distinctly different parts of the Fermi surface \cite{PhysRevLett.3.552}. The interest in this phenomenon and the emergent new physics was invigorated after the experimental discovery of two-gap superconductivity in bulk MgB$_2$ in 2001 \cite{Nagamatsu2001}. MgB$_2$ consists of planes of boron in a honeycomb lattice alternated by planes of Mg-atoms sitting above the centers of the honeycomb tiles. It is therefore akin to intercalated graphite \cite{Weller2005}, with Mg in the role of the dopant. In MgB$_2$, in-plane $\sigma$-bonds coexist with out-of-plane $\pi$-bonds, and separately give rise to two superconducting gaps for bulk MgB$_2$: the stronger $\sigma$-gap $\Delta_{\sigma}(0)\sim 7$ meV and the weaker $\pi$-gap, $\Delta_{\pi}(0) \sim 2 - 3$ meV \cite{0953-2048-16-2-305,Choi2002,PhysRevB.91.214519,PhysRevB.87.024505,PhysRevB.92.054516}.

Competition and coupling between the multiple condensates in a multi-gap superconductor can lead to rich new physics \cite{0953-2048-28-6-060201}. In that sense, one expects superconductors with three or more gaps to be far more exciting than the two-gap ones, due to additional competing effects and possible quantum frustration between the condensates \cite{PhysRevB.81.134522}. To date discovered effects specific to multi-gap superconductors include novel vortical and skyrmionic states \cite{PhysRevLett.89.067001,PhysRevLett.107.197001}, giant-paramagnetic response \cite{rogerio}, hidden criticality \cite{PhysRevLett.108.207002}, and time-reversal symmetry breaking \cite{PhysRevB.81.134522,PhysRevB.87.134510}, to name a few. A major roadblock for the experimental confirmation of these predictions is the lack of distinctly multi-gap (beyond two-gap) superconductors. In recent years two such materials were proposed theoretically by Gross and coworkers, using density functional theory for superconductors \cite{PhysRevLett.115.097002}. One is molecular hydrogen, which under very high pressure develops three superconducting gaps on different Fermi sheets \cite{PhysRevLett.100.257001}. However, due to anisotropy two of the gaps strongly overlap. The other material is CaBeSi, a MgB$_2$-like compound in which splitting of the $\pi$-bands was predicted to give rise to three-gap superconductivity \cite{PhysRevB.79.104503}, but with impractically low $T_{\mathrm{c}} \cong 0.4$ K.

Here, we follow a different route, namely that of atomically-thin instead of bulk superconductors. Recently, owing to immense experimental progress \cite{0953-2048-30-1-013003,0953-2048-30-1-013002}, superconductivity was realized down to monolayer thickness in several materials -- ranging from electron-phonon-based superconductors, such as In and Pb \cite{Qin2009,Zhang2010}, NbSe$_2$ \cite{doi:10.1021/acs.nanolett.5b00648,Ugeda2016,Xi2016} and doped graphene \cite{Profeta2012,2053-1583-1-2-021005,Ludbrook22092015,Kanetani27112012,Chapman2016,2053-1583-3-4-045003}, to materials with non-conventional coupling mechanisms, such as La$_{2-x}$Sr$_x$CuO$_4$ \cite{Bollinger2011} and FeSe \cite{Ge2015}. The promise for extremely low power, ultra-lightweight and ultra-sensitive electronic devices warrants further progress in ultrathin superconductivity \cite{Golod2015,Najafi2015,Lowell2016}. Quantum confinement in the vertical direction generally separates subbands in ultrathin films, innating multi-band and thereby potentially multi-gap superconductivity \cite{PhysRevB.87.064510}. We here note an additional, natural connection between two-dimensional and multi-gap superconductors, much less explored to date: surface states can equally host new superconducting gaps without equivalent in the bulk material. 

In this paper, we start from the known bulk two-gap superconductor MgB$_2$, and show how the gap spectrum changes at the thinnest limit. It was predicted that, albeit not being the thermodynamic ground state, such structures are mechanically stable and could be grown owing to kinetic barriers \cite{PhysRevB.80.134113}, such that few-monolayer MgB$_2$ has already been synthesized experimentally on a Mg-substrate \cite{Cepek2004}. Using a combination of first-principles calculations and anisotropic Eliashberg theory, we reveal a major influence of an emerging surface state on superconductivity in these ultrathin films. This contribution hybridizes with those of the $\sigma$- and $\pi$-bands in a highly nontrivial manner, changing the multi-gap physics with every additional monolayer. This finally leads to pure three-gap superconductivity in one-monolayer MgB$_2$, retained up to a high critical temperature of 20 K (highest among monolayer superconductors without coupling to a substrate). This superconductivity originating from the surface state could not be detected by a previous study of few-monolayer MgB$_2$ based on the tight-binding formalism, in which surface states (electronic as well as vibrational) were completely omitted \cite{PhysRevB.74.094501}. We further demonstrate that this three-gap superconductivity remains robust even under strain, where tensile strain of just $\sim$ 4\% boosts $T_{\mathrm{c}}$ to above 50 K. Such small strain was previously found to increase $T_{\mathrm{c}}$ in bulk MgB$_2$ by at most 10\% \cite{PhysRevLett.93.147006,PhysRevB.73.024509}, or nearly not at all in both electron- \cite{0295-5075-108-6-67005} and hole-doped \cite{PhysRevLett.111.196802,C5NR07755A} graphene (only strain beyond 5\% is predicted to have significant influence there). Considering that such straining can be conveniently realized by growing the monolayer MgB$_2$ on substrates with a somewhat larger lattice constant (e.g., Si$_{1+x}$C$_{1-x}$ or Al$_x$Ga$_{1-x}$N alloys, with a lattice constant tunable by $x$) \cite{PhysRevB.73.024509}, we expect our results to be of immediate experimental relevance.

\begin{figure*}[ht]
\centering
\includegraphics[width=1\linewidth]{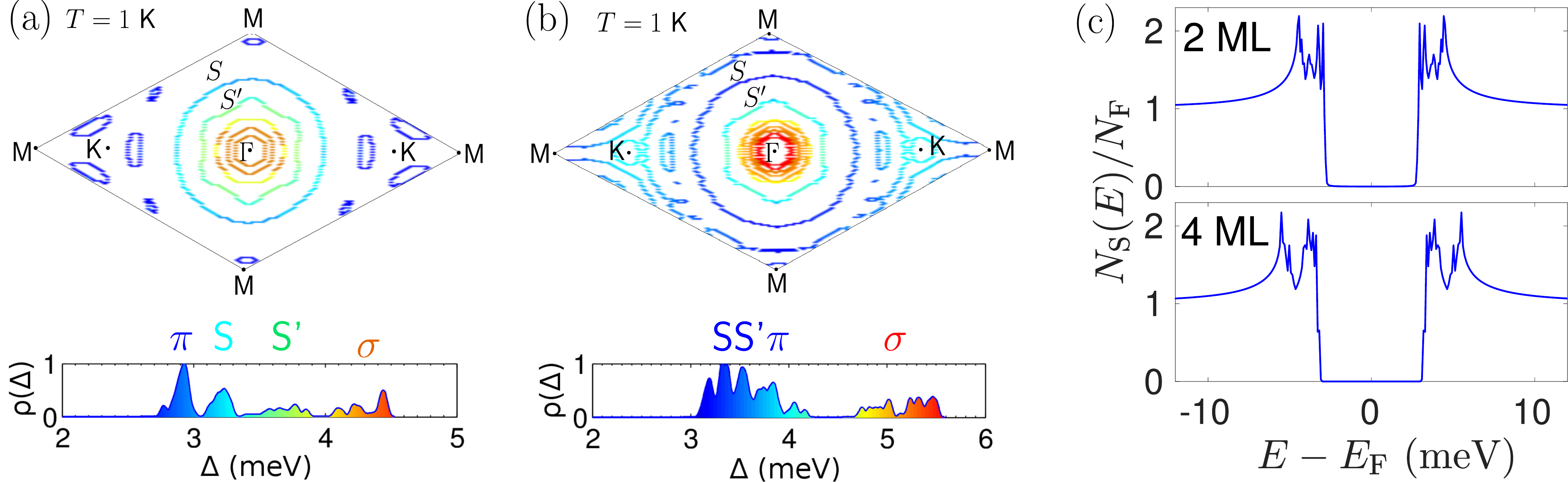}
\caption{(Color online) (a) and (b) The distribution of the superconducting gap spectrum of 2-ML and 4-ML MgB$_2$, respectively, on the Fermi surface, calculated from anisotropic Eliashberg theory with \textit{ab initio} input. Both are anisotropic two-gap superconductors, with surface condensates \textit{S} and \textit{S'} hybridized with the $\pi$ condensate. (c) The density of states in the superconducting state for 2 and 4 MLs, calculated at $T=1$ K, showing the overall two gap-nature as well as the anisotropy of the gap spectrum. The critical temperatures found for 2 and 4 MLs MgB$_2$ are 23 K and 27 K respectively.}
\label{fig:fig2}
\end{figure*}
\begin{figure}[ht]
\centering
\includegraphics[width=0.8\linewidth]{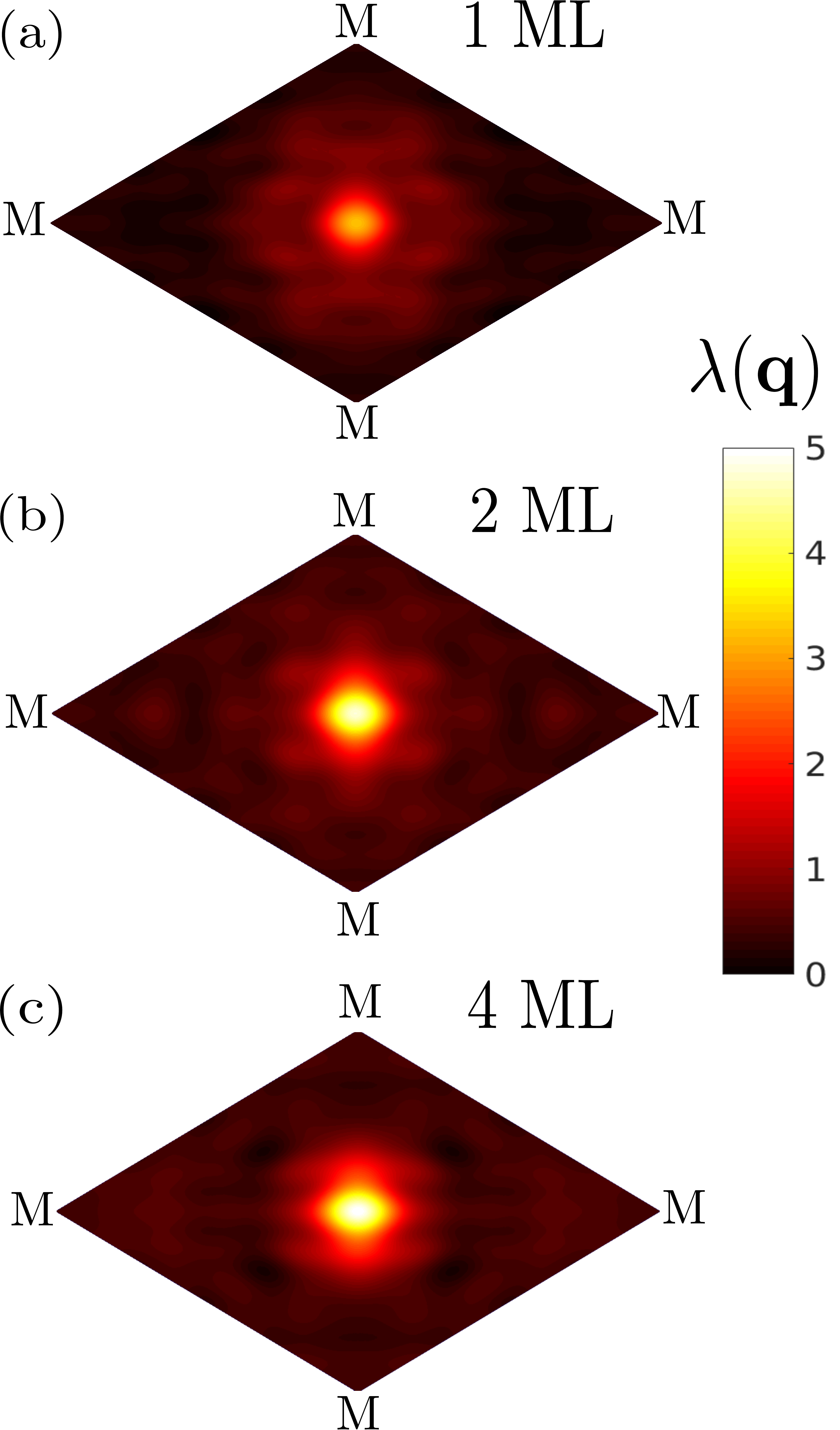}
\caption{(Color online) The overall \textit{e-ph} coupling $\lambda(\textbf{q})=\sum_{\nu} \lambda_{\nu}(\textbf{q})$ (i.e., summed over all phonon nodes) as a function of phonon wave vectors $\textbf{q}$ for (a) 1-ML, (b) 2-ML and (c) 4-ML MgB$_2$.} 
\label{fig:fig3}
\end{figure}

\section{Monolayer MgB$_2$}

Our investigation starts from first-principles calculations (using ABINIT \cite{Gonze20092582,suppl_mat}) of one monolayer (ML) of MgB$_2$. It consists of one Mg- and one B-layer, the latter in a honeycomb lattice, and thus structurally similar to doped graphene. The resulting Fermi surface is shown in Fig.~\ref{fig:fig1}(a). It consists of two $\sigma$-bands (around $\Gamma$), a $\pi$-band (around K), and a surface band \textit{S}. While, as we mentioned above, the former two are also present in bulk MgB$_2$, the surface band originates from the Mg-plane facing vacuum. It is thus characteristic of two-dimensional forms of MgB$_2$ and has predominant Mg-$p$ character, as opposed to the B-$p$ character of the other bands. Next, we calculated the electron-phonon (\textit{e-ph}) coupling in 1-ML MgB$_2$ from first principles, employing density functional perturbation theory (DFPT) \cite{PhysRevB.54.16487,Gonze20092582}. With this input, the anisotropic Eliashberg equations (i.e., taking into account the full spatial dependence) were solved self-consistently \cite{PhysRevB.92.054516,PhysRevB.94.144506,suppl_mat}.

We describe the Coulomb repulsion with $\mu^{*}=0.13$, yielding correct $T_{\mathrm{c}}$ for bulk MgB$_2$. This value is also in line with previously established values \cite{Choi2002,PhysRevB.70.104522}. The Coulomb pseudopotential is not expected to change drastically in the 2D limit, owing to the layered structure of MgB$_2$. Namely, superconductivity of the dominant $\sigma$-bands is quasi-two-dimensional even in bulk MgB$_2$, so the same is expected for the screening.  
 
In Fig.~\ref{fig:fig1}(a) we show the resulting superconducting gap spectrum on the Fermi surface, $\Delta(\textbf{k}_{\mathrm{F}},T)$, at $T=1$ K, as well as the distribution of the gap, $\rho(\Delta)$. This result shows that 1-ML MgB$_2$ is a distinctly three-gap superconductor, with separate gaps opening on the $\sigma$-, $\pi$- and \textit{S}-bands. The gap amplitudes are about half of those of bulk MgB$_2$, with Fermi surface averages at zero temperature of $\langle \Delta_{\sigma}(0) \rangle = 3.3$ meV, $\langle \Delta_{S}(0) \rangle = 2.7$ meV and $\langle \Delta_{\pi}(0) \rangle = 1.4$ meV. The critical temperature of $T_{\mathrm{c}}=20$ K, compared to the bulk $T_{\mathrm{c}} \cong 39$ K \cite{0953-2048-16-2-305,Choi2002,PhysRevB.91.214519,PhysRevB.87.024505,PhysRevB.92.054516}, follows the same trend. 

To corroborate further the predicted three-gap superconductivity in 1-ML MgB$_2$, we calculated the density of states (DOS) in the superconducting state $N_{\mathrm{S}}$, using Eliashberg relations \cite{PhysRevB.92.054516,suppl_mat}. The result displayed in Fig.~\ref{fig:fig1}(b) shows that $N_{\mathrm{S}}$ for 1-ML MgB$_2$ consists of three distinct and narrow peaks, corresponding to the three superconducting gaps. As $N_{\mathrm{S}}$ determines the superconducting tunneling properties, the predicted three-gap superconductivity can be verified with low-temperature scanning tunneling spectroscopy \cite{0953-2048-16-2-305}.

Last but not least, we show that three-gap superconductivity in 1-ML MgB$_2$ is very robust with temperature. Fig.~\ref{fig:fig1}(c) displays the calculated temperature evolution of the superconducting gap spectrum, proving that the three superconducting gaps are well separated up to 18 K, very close to $T_{\mathrm{c}} = 20$ K.

\section{Evolution with added monolayers}

To provide a deeper understanding of the origin of three-gap superconductivity in 1-ML MgB$_2$, we studied what changes when adding monolayers to the system, considering in particular 2- and 4-ML thick MgB$_2$. The superconducting gap spectra, obtained using anisotropic Eliashberg theory, are displayed in Fig.~\ref{fig:fig2} (a) and (b). One observes in Fig.~\ref{fig:fig2}(a) that a hexagonal band lying between the \textit{S}-band and the $\sigma$-bands develops an additional gap in 2-ML MgB$_2$. This band is a split-off band of the $\sigma$-bands (with B-$p$ character), indicated with \textit{S'} as it originates from a surface state of the free B-surface. The superconducting gap opening on band \textit{S'} is weakly linked to the gaps opening on the $\pi$- and \textit{S}-bands, but (barely) separate from the gap on the $\sigma$-bands, making 2-ML MgB$_2$ an anisotropic two-gap (nearly single-gap) superconductor. In 4-ML MgB$_2$ we find a higher degree of hybridization between the $\pi$-, \textit{S}- and \textit{S'}-condensates, forming an anisotropic gap clearly separated from the $\sigma$-gap. In Fig.~\ref{fig:fig2}(c) we show the corresponding DOS in the superconducting state. For 2-ML MgB$_2$, $N_{\mathrm{S}}$ clearly reflects the anisotropy of the gap spectrum, while for 4-ML MgB$_2$ $N_{\mathrm{S}}$ consists of two broader peaks, resulting from the strong hybridization between the condensates. The critical temperatures we obtained from the solution of the anisotropic Eliashberg equations are larger than that of 1-ML MgB$_2$, namely 23 K and 27 K for 2-ML and 4-ML MgB$_2$ respectively (still well below the bulk value of 39 K \footnote{We note here that this result is different from that obtained in Ref.~\citenum{MORSHEDLOO20151} for 2-ML MgB$_2$, where $T_{\mathrm{c}}$ was found to exceed the bulk value. The difference can be traced back to the unreasonably low Coulomb pseudopotential used in this work, to compensate the lack of multi-band effects in their isotropic Eliashberg approach.}).

The transition from three-gap superconductivity in ML MgB$_2$ to anisotropic two-gap superconductivity and 2-ML and 4-ML MgB$_2$ can be explained by means of the \textit{e-ph} coupling field shown in Fig.~\ref{fig:fig3}. In all cases, the \textit{e-ph} coupling peaks for phonon wave vectors $\textbf{q} \simeq 0$ (i.e., $\Gamma$), which promotes intraband coupling, giving rise to separate condensates on different sheets. However, in Fig.~\ref{fig:fig3} one observes also a clear evolution towards stronger coupling at non-zero wave vectors going from a ML to thicker structures. These emerging coupling channels enable scattering between different sheets, notably between the close-lying \textit{S}, \textit{S'} and $\pi$-bands. This leads to the hybridization between the corresponding condensates shown in Fig.~\ref{fig:fig2}.

Our results show thus a drastic change from the distinctly three-gap superconductivity in single ML MgB$_2$ to very anisotropic two-gap superconductivity by addition of even a single monolayer. Bearing in mind that the superconducting gap opening on the surface band in very thick MgB$_2$ films was found experimentally to be nearly degenerate with the gap on the $\sigma$ band \cite{Souma03}, we expect further rich behavior of the gap spectrum as the MgB$_2$ film is made progressively thicker beyond 4 MLs. Besides accompanying fundamental physics, this strong variation of the gap structure with the number of MLs opens perspectives for nano-engineered superconducting junctions using one single material with spatially varied thickness on the atomic scale. Such local control of thickness is readily available for, e.g., Pb films \cite{0953-2048-30-1-013003,0953-2048-30-1-013002}.
\begin{figure}[ht]
\centering
\includegraphics[width=1\linewidth]{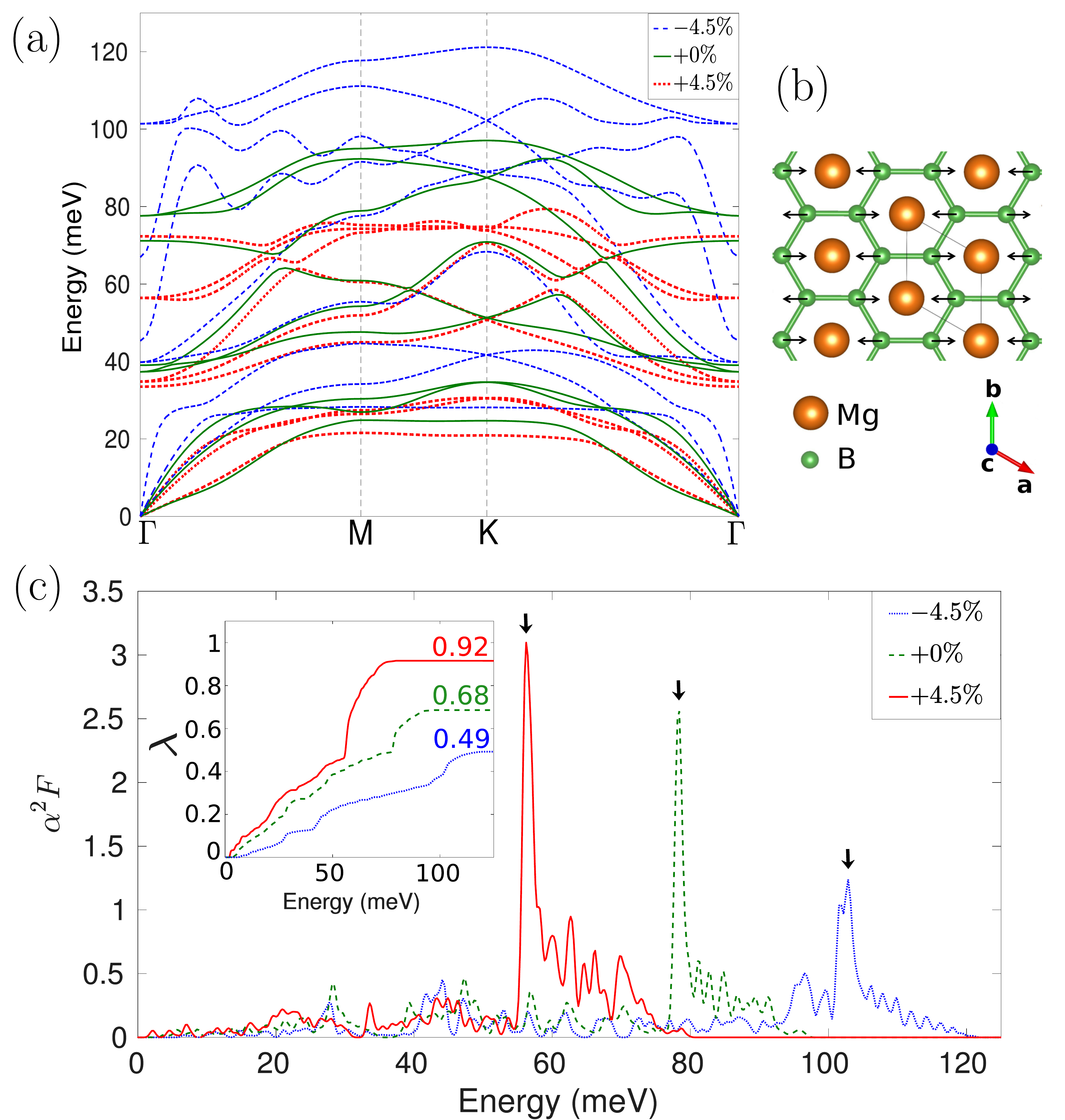}
\caption{(Color online) Phonons and electron-phonon coupling of biaxially strained 1-ML MgB$_2$ calculated using DFPT. (a) The phonon dispersion for strains of $-4.5\%$, $+0\%$ and $+4.5\%$. Increasing strain leads to lower phonon frequencies. (b) The E$_{2g}$ phonon mode of the B-atoms that gives the strongest contribution to the electron-phonon coupling. (c) The isotropic Eliashberg function under different strains, $\alpha^2F(\omega)=\langle\langle\alpha^2F({\bf k\, k'},\omega)\rangle_{{\bf k}'_{\mathrm{F}}}\rangle_{{\bf k}_{\mathrm{F}}}$ (i.e., the double Fermi surface average). The peaks originating from the E$_{2g}$ mode are indicated by arrows. The resulting electron-phonon coupling $\lambda$ is shown as inset.}
\label{fig:fig4}
\end{figure}
\begin{figure}[ht]
\centering
\includegraphics[width=0.95\linewidth]{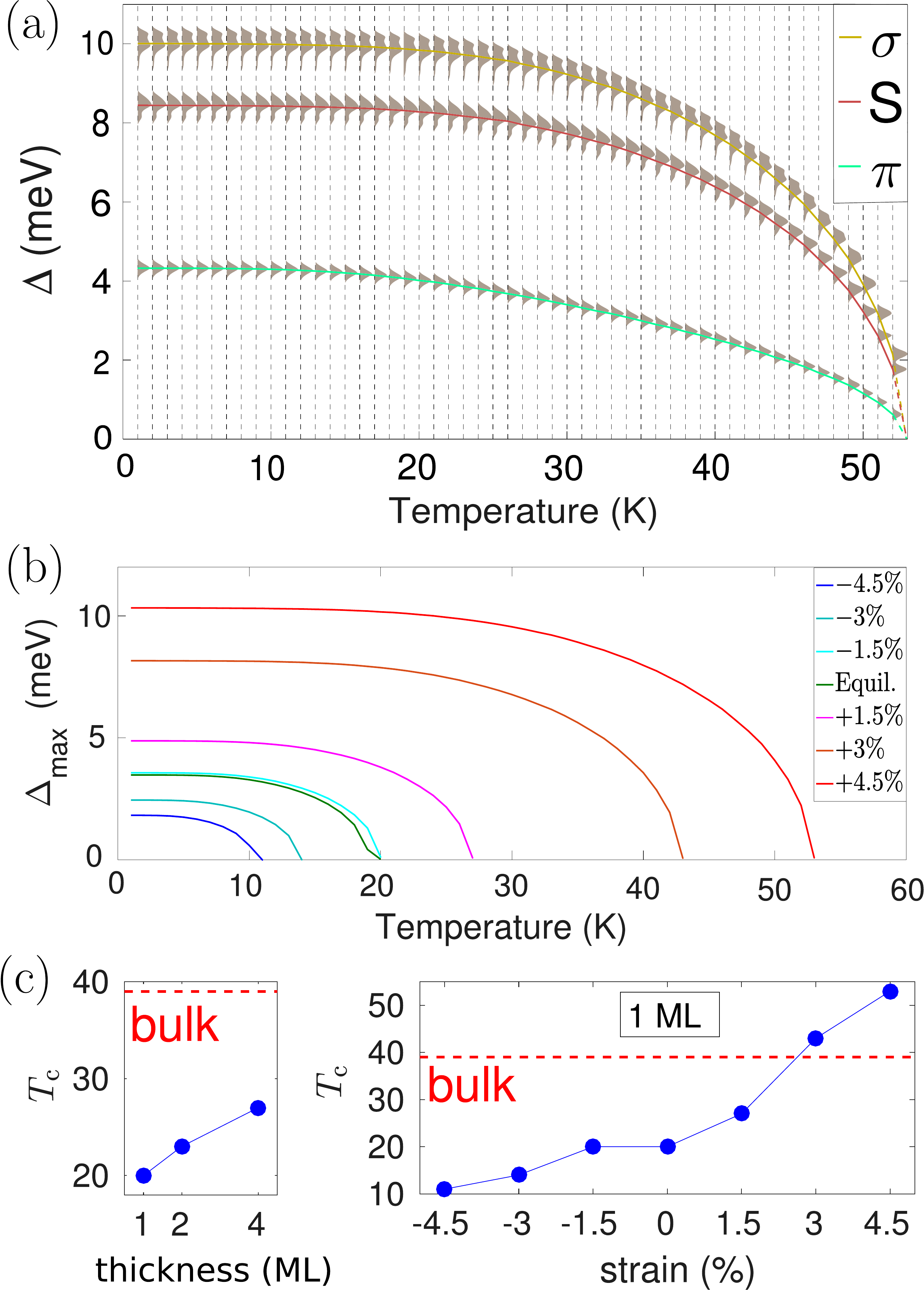}
\caption{(Color online) The superconducting spectrum of a biaxially strained 1-ML MgB$_2$. (a) The distribution of the superconducting gap for $+4.5\%$ tensile strain as a function of temperature, displaying the same three gaps ($\pi$, \textit{S} and $\sigma$) as in the unstrained case (Fig.~\ref{fig:fig1}). The calculation shows an enhancement of the critical temperature to $T_{\mathrm{c}}=53$ K. (b) The maximum value of the superconducting gap, $\Delta_{\mathrm{max}}$, as a function of temperature and strain. Superconductivity depletes upon compression and is strongly boosted with tensile strain. (c) $T_{\mathrm{c}}$ as a function of the film thickness, and as a function of strain for a 1-ML MgB$_2$. The bulk value, $T_{\mathrm{c}}=39$ K, is shown for comparison.}
\label{fig:fig5}
\end{figure}

\section{Strained monolayer MgB$_2$}

In experiments, the preferred growth method of atomically thin MgB$_2$ is epitaxial growth on a substrate \cite{Cepek2004}. Due to the ever-present lattice mismatch in that case, we consider the effect of strain on the three-gap superconductivity predicted here. We concentrate on biaxial strain applied with respect to the in-plane cell parameter, namely the Mg-Mg distance with equilibrium value $a=3.04$ \AA. In Fig.~\ref{fig:fig4}(a) we compare the equilibrium phonon band structure of 1-ML MgB$_2$ with the cases of $-4.5$\% compressive strain and $+4.5$\% tensile strain. In the tensile case, interatomic charge densities get depleted as the distances between atoms increase. Consequently, the interatomic bonds become less stiff, resulting in a decrease of phonon frequencies. In the compressive case, the exact opposite occurs. In Fig.~\ref{fig:fig4}(b) we show the E$_{2g}$ phonon mode of the B-atoms, which is the mode harbouring the strongest \textit{e-ph} coupling in 1-ML MgB$_2$. As such, this mode dominates the Eliashberg function $\alpha^2F$, shown in Fig.~\ref{fig:fig4}(c). The peaks in $\alpha^2F$ due to the E$_{2g}$ mode (indicated by arrows) are stronger and more pronounced in 1-ML MgB$_2$ compared with bulk MgB$_2$ \cite{PhysRevB.92.054516}, in particular in equilibrium and under tensile strain. The shift to lower energy (following the general trend for the phonons) and amplification of this peak due to tensile strain lead to a significant enhancement of the \textit{e-ph} coupling, as shown in the inset of Fig.~\ref{fig:fig4}(c). As follows from the above discussion, it is a general principle that tensile strain lowers the energy of the phonon modes, resulting in enhanced \textit{e-ph} coupling, since $\lambda=2\int_0^\infty \mathrm{d}\omega\omega^{-1}\alpha^2F(\omega)$ is weighted by $\omega^{-1}$ \cite{Grimvall}. However, the effect is particularly strong in 1-ML MgB$_2$ due to the occurrence of the E$_{2g}$ phonon mode, which not only goes down in energy but also develops stronger intrinsic coupling to electrons, as follows from the evolution of the Eliashberg function shown in Fig.~\ref{fig:fig4}(c). A similar trend in the \textit{e-ph} coupling under the influence of strain has been found in both electron- and hole-doped graphene \cite{0295-5075-108-6-67005,PhysRevLett.111.196802,C5NR07755A}, although much less pronounced.

With this first-principles input for strained 1-ML MgB$_2$, we solved again the anisotropic Eliashberg equations. We found that the Fermi surface is almost unaltered  w.r.t.~that shown in Fig.~\ref{fig:fig1}(a), in the studied range of straining of $-4.5$\% to $+4.5$\%. This, in combination with the robust coupling to the E$_{2g}$ mode, leads to three-gap superconductivity in ML MgB$_2$ being conserved under all strains considered here \footnote{We note that for compressive strains exceeding $-1.5$\% $\sigma$- and \textit{S}-gaps become hybridized, albeit their contributions can still be distinguished. Their partial overlap is not due to new physics -- it is provoked by a general depletion of the superconducting gap values, forcing the gaps closer together.}. In Fig.~\ref{fig:fig5}(a) we show the temperature evolution of the gap spectrum of 1-ML MgB$_2$ subject to tensile strain of $+4.5$ \%, proving the robustness of the three-gap superconductivity even under a considerable amount of strain. Owing to the enhanced \textit{e-ph} coupling [cf.~Fig.~\ref{fig:fig4}(c)] the superconducting gaps are much larger than in the equilibrium case. For $+4.5$ \% strain, the average gaps amount to $\langle \Delta_{\sigma}(0) \rangle = 10.0$ meV, $\langle \Delta_{S}(0) \rangle = 8.4 $ meV and $\langle \Delta_{\pi}(0) \rangle =4.3$ meV, with a corresponding critical temperature as high as $T_{\mathrm{c}} = 53$ K. In Fig.~\ref{fig:fig5}(b) we show the temperature evolution of the maximum ($\sigma$) gap value, comparatively for different strains. It reveals that upon compression, superconductivity is greatly suppressed ($T_{\mathrm{c}}$ drops to 11 K for $-4.5$\% strain), while it is strongly boosted when the ML is subject to tensile strain. The changes are particularly drastic for such limited amounts of strain, in comparison to, e.g., superconducting doped graphene \cite{0295-5075-108-6-67005,PhysRevLett.111.196802,C5NR07755A}. In Fig.~\ref{fig:fig4}(c) we show the evolution of $T_{\mathrm{c}}$ with the number of monolayers and with strain. It is apparent that the effect of strain on superconductivity is stronger, with a ML strained at $+3$\% already surpassing bulk MgB$_2$ as to its $T_{\mathrm{c}}$. A major difference between both manipulations we considered is that strain preserves the three-gap superconductivity of monolayer MgB$_2$, while increasing thickness strongly changes the gap spectrum with every added monolayer, as shown in Fig.~\ref{fig:fig2}.

\section{Conclusion}

In summary, we presented the formation and evolution of three-gap superconductivity in few-monolayer MgB$_2$, by solving the anisotropic Eliashberg equations with full \textit{ab initio} input. We showed that the electronic surface band, originating from the free Mg-surface, plays a major role in ultrathin MgB$_2$, and hosts a third superconducting gap that coexists with the bulk-like $\pi$- and $\sigma$-gaps. These gaps are distinctly separate in 1-monolayer MgB$_2$, where the resulting three pronounced peaks in the superconducting tunneling spectrum provide a clear signature for experimental validation of our prediction. The shown three-gap superconductivity is moreover very robust with temperature, persisting even close to the critical temperature of 20 K. With only $\sim 4$\% tensile strain, \textit{e-ph} coupling is greatly enhanced and superconductivity is boosted to temperatures beyond $50$ K. As more monolayers are added to the film, different condensates hybridize, changing the multi-gap spectrum drastically with every added monolayer. Our investigation therefore establishes atomically thin MgB$_2$ as a unique system to explore tunability of high-$T_\mathrm{c}$, multi-gap superconductivity, and its possible applications in ultrathin cryogenic electronics engineered by strain and atomically controlled thickness. 

\begin{acknowledgments}
\noindent This work was supported by TOPBOF-UAntwerp, Research Foundation-Flanders (FWO), the Swedish Research Council (VR) and the R{\"o}ntgen-{\AA}ngstr{\"o}m Cluster. The first-principles calculations have been carried out on the HPC infrastructure of the University of Antwerp (CalcUA), a division of the Flemish Supercomputer Centre (VSC), supported financially by the Hercules foundation and the Flemish Government (EWI Department). Eliashberg theory calculations were supported through the Swedish National Infrastructure for Computing (SNIC). 
\end{acknowledgments}

\section*{Appendix}

\begin{appendix}

\section{Density functional (perturbation) theory calculations}

Our density functional theory (DFT) calculations make use of the Perdew-Burke-Ernzerhof (PBE) functional implemented within a planewave basis in the ABINIT code \cite{Gonze20092582}. Electron-ion interactions are treated using norm-conserving Vanderbilt pseudopotentials \cite{PhysRevB.88.085117}, taking into account Mg-2$s^2$2$p^6$3$s^2$ and B-2$s^2$2$p^1$ as valence electrons. An energy cutoff of 60 Ha for the planewave basis was used, to achieve convergence of the total energy below 1 meV per atom. In order to simulate the atomically thin films, we used unit cells that include 25 \AA~of vacuum. A dense $22 \times 22 \times 1$ $ \Gamma$-centered Monkhorst-Pack \textbf{k}-point grid is used for an accurate description of the Fermi surfaces. The lattice parameters were obtained using a conjugate-gradient algorithm so that forces on each atom were minimized below 1 meV/\AA. Strain was implemented by changing the in-plane lattice parameter w.r.t.~the equilibrium value thus obtained.\\
\indent  To calculate phonon dispersions and electron-phonon coupling, density functional perturbation theory (DFPT) calculations were carried out, also within the framework of ABINIT. The total number of perturbations due to atomic displacements to be treated (in other words, the number of phonon branches) amounts to $3\cdot N_{\mathrm{atoms}}$, ranging from 9 for a ML to 36 for 4 MLs. Thus, the phonon spectrum and electron-phonon coupling coefficients, matrix elements of the perturbative part of the Hamiltonian \cite{PhysRevB.54.16487}, are obtained. We carried out the DFPT calculations on a $22 \times 22 \times 1$ electronic $\textbf{k}$-point grid and a $11 \times 11 \times 1~\textbf{q}$-point grid (a subgrid of the $\textbf{k}$-point grid) as phonon wave vectors.

\section{Fully anisotropic Eliashberg theory calculations}

In order to describe superconductivity of MgB$_2$  on an \textit{ab initio} level, we solve self-consistently the coupled anisotropic Eliashberg equations \cite{Choi2002},
\begin{align} 
Z\ph_{{\bf k},n}&= 1+ \frac{\pi T}{\omega_n}\sum_{{\bf k'},n'}\frac{\delta(\xi\ph_{\bf k'})}{N_F}\lambda({\bf kk'},nn')  \nonumber \\  \label{el1} & \times \frac{\omega_{n'}}{\sqrt{\omega_{n'}^2 + \Delta^2_{{\bf k'},n}}}  \\ 
\Delta\ph_{{\bf k},n}Z\ph_{{\bf k},n}&=\pi T\sum_{{\bf k'},n'}\frac{\delta(\xi\ph_{\bf k'})}{N_F}\left[\lambda({\bf kk'},nn') - \mu^*(\omega_c)\right] \nonumber \\  \label{el2} & \times \frac{\Delta\ph_{{\bf k'},n'}}{\sqrt{\omega_{n'}^2 + \Delta^2_{{\bf k},n}}}
\end{align}
using the \textit{ab initio} calculated electron band structure contained in $\xi\ph_{\bf k}$ and phonon and electron-phonon coupling contained in $\lambda({\bf kk'},nn')$. In the above, T is temperature, $\omega_n=\pi T(2n+1)$ are fermion Matsubara frequencies, $Z\ph_{{\bf k},n}$ is the mass renormalization function, $\Delta\ph_{{\bf k},n}$ describes anisotropic even-frequency spin singlet superconductivity, $N_{\mathrm{F}}$ is the electronic density of states at the Fermi level and $\mu^*(\omega_c)$ is the Anderson-Morel Coulomb pseudopotential which comes with a cut-off $\omega_c$. The momentum dependent electron-phonon coupling is
\begin{align}
\lambda({\bf k-k'},n-n')&= \nonumber \\ 
\int_0^\infty d\omega \, \alpha^2F({\bf k\, k'},\omega) & \frac{2\omega}{\left(\omega_n-\omega_{n'}\right)^2+\omega^2},
\end{align}
with the momentum dependent Eliashberg function
\begin{eqnarray}
\alpha^2F({\bf k\, k'},\omega)=N_F\sum_{\nu} |g^\nu_{\bf q}|^2 \delta(\omega-\omega\ph_{{\bf q}\nu}),
\end{eqnarray}
where ${\bf q}={\bf k}-{\bf k'}$ and where $g^\nu_{\bf q}$ and $\omega\ph_{{\bf q}\nu}$ are the phonon branch-resolved electron-phonon scattering matrix elements and phonon frequencies, respectively. From the above, one can obtain the isotropic Eliashberg function as 
\begin{eqnarray}
\alpha^2F(\omega)=\langle\langle\alpha^2F({\bf k\, k'},\omega)\rangle_{{\bf k}'_{\mathrm{F}}}\rangle_{{\bf k}_{\mathrm{F}}},
\end{eqnarray}
where $\langle\ldots\rangle_{{\bf k}_{\mathrm{F}}}=\frac{1}{N_{\mathrm{F}}}\sum_{\bf k}\delta(\xi\ph_{\bf k})\left(\ldots\right)$ is the Fermi surface average.

The quasiparticle density of states that is proportional to single-particle tunneling measurements, is given by
\begin{eqnarray}
N_{\mathrm{S}}(\Omega)\propto\sum_{\bf k}A({\bf k},\Omega)\approx N_{\mathrm{F}}\Big\langle \int_{-\infty}^{\infty}d\xi A_{\bf k}(\xi,\Omega)\Big\rangle_{{\bf k}_{\mathrm{F}}}
\end{eqnarray}
with the spectral function,
\begin{eqnarray}
A({\bf k},\Omega)=-\frac{1}{\pi}\textrm{Im}\left[\hat{G}_R\ph({\bf k},\Omega)\right]_{11}
\end{eqnarray}
where $\left[\hat{G}_R\ph({\bf k},\Omega)\right]_{11}$ is the (11) element of the retarded matrix Green's function, obtained after analytic continuation of the full matrix Green's function,
\begin{eqnarray}\label{Gfun}
\hat{G}\ph_{{\bf k},n}=\left[i\omega_n Z\ph_{{\bf k},n}\hat{\rho}_0-\xi\ph_{\bf k}\hat{\rho}_3-\Delta\ph_{{\bf k},n}\hat{\rho}_1\right]^{-1}.
\end{eqnarray}
The coupled equations (\ref{el1}--\ref{el2}), supplemented by the electron and phonon band structure and the electron-phonon coupling, calculated by first principles, were solved self-consistently in Matsubara space and the converged solutions were then analytically continued to real frequencies. In order to ensure a good accuracy, we imposed a strict convergence criterion of $\frac{x_n-x_{n-1}}{x_n}<10^{-6}$ and allowed up to 1000 iteration cycles. In all the calculations presented here we set $\mu^*(\omega_c)=0.13$ for the Coulomb pseudopotential with a cut-off frequency $\omega_c>0.5$ eV. We have also checked that $\omega_c$ is sufficiently large and that results are well converged with this cut-off. The analytic continuation was performed numerically by employing  the high-accuracy Pad\'e scheme based on symbolic computation \cite{PhysRevB.61.5147,PhysRevB.92.054516} with a chosen precision of 250 decimal digits. After this procedure, we calculate the retarded momentum dependent Green's function, the tunneling spectra and the superconducting gap-edge.

\end{appendix}

\bibliography{biblio}

\begin{thebibliography}{57}%
\makeatletter
\providecommand \@ifxundefined [1]{%
 \@ifx{#1\undefined}
}%
\providecommand \@ifnum [1]{%
 \ifnum #1\expandafter \@firstoftwo
 \else \expandafter \@secondoftwo
 \fi
}%
\providecommand \@ifx [1]{%
 \ifx #1\expandafter \@firstoftwo
 \else \expandafter \@secondoftwo
 \fi
}%
\providecommand \natexlab [1]{#1}%
\providecommand \enquote  [1]{``#1''}%
\providecommand \bibnamefont  [1]{#1}%
\providecommand \bibfnamefont [1]{#1}%
\providecommand \citenamefont [1]{#1}%
\providecommand \href@noop [0]{\@secondoftwo}%
\providecommand \href [0]{\begingroup \@sanitize@url \@href}%
\providecommand \@href[1]{\@@startlink{#1}\@@href}%
\providecommand \@@href[1]{\endgroup#1\@@endlink}%
\providecommand \@sanitize@url [0]{\catcode `\\12\catcode `\$12\catcode
  `\&12\catcode `\#12\catcode `\^12\catcode `\_12\catcode `\%12\relax}%
\providecommand \@@startlink[1]{}%
\providecommand \@@endlink[0]{}%
\providecommand \url  [0]{\begingroup\@sanitize@url \@url }%
\providecommand \@url [1]{\endgroup\@href {#1}{\urlprefix }}%
\providecommand \urlprefix  [0]{URL }%
\providecommand \Eprint [0]{\href }%
\providecommand \doibase [0]{http://dx.doi.org/}%
\providecommand \selectlanguage [0]{\@gobble}%
\providecommand \bibinfo  [0]{\@secondoftwo}%
\providecommand \bibfield  [0]{\@secondoftwo}%
\providecommand \translation [1]{[#1]}%
\providecommand \BibitemOpen [0]{}%
\providecommand \bibitemStop [0]{}%
\providecommand \bibitemNoStop [0]{.\EOS\space}%
\providecommand \EOS [0]{\spacefactor3000\relax}%
\providecommand \BibitemShut  [1]{\csname bibitem#1\endcsname}%
\let\auto@bib@innerbib\@empty
\bibitem [{\citenamefont {Suhl}\ \emph {et~al.}(1959)\citenamefont {Suhl},
  \citenamefont {Matthias},\ and\ \citenamefont {Walker}}]{PhysRevLett.3.552}%
  \BibitemOpen
  \bibfield  {author} {\bibinfo {author} {\bibfnamefont {H.}~\bibnamefont
  {Suhl}}, \bibinfo {author} {\bibfnamefont {B.~T.}\ \bibnamefont {Matthias}},
  \ and\ \bibinfo {author} {\bibfnamefont {L.~R.}\ \bibnamefont {Walker}},\
  }\href {\doibase 10.1103/PhysRevLett.3.552} {\bibfield  {journal} {\bibinfo
  {journal} {Phys. Rev. Lett.}\ }\textbf {\bibinfo {volume} {3}},\ \bibinfo
  {pages} {552} (\bibinfo {year} {1959})}\BibitemShut {NoStop}%
\bibitem [{\citenamefont {Nagamatsu}\ \emph {et~al.}(2001)\citenamefont
  {Nagamatsu}, \citenamefont {Nakagawa}, \citenamefont {Muranaka},
  \citenamefont {Zenitani},\ and\ \citenamefont {Akimitsu}}]{Nagamatsu2001}%
  \BibitemOpen
  \bibfield  {author} {\bibinfo {author} {\bibfnamefont {J.}~\bibnamefont
  {Nagamatsu}}, \bibinfo {author} {\bibfnamefont {N.}~\bibnamefont {Nakagawa}},
  \bibinfo {author} {\bibfnamefont {T.}~\bibnamefont {Muranaka}}, \bibinfo
  {author} {\bibfnamefont {Y.}~\bibnamefont {Zenitani}}, \ and\ \bibinfo
  {author} {\bibfnamefont {J.}~\bibnamefont {Akimitsu}},\ }\href {\doibase
  10.1038/35065039} {\bibfield  {journal} {\bibinfo  {journal} {Nature}\
  }\textbf {\bibinfo {volume} {410}},\ \bibinfo {pages} {63} (\bibinfo {year}
  {2001})}\BibitemShut {NoStop}%
\bibitem [{\citenamefont {Weller}\ \emph {et~al.}(2005)\citenamefont {Weller},
  \citenamefont {Ellerby}, \citenamefont {Saxena}, \citenamefont {Smith},\ and\
  \citenamefont {Skipper}}]{Weller2005}%
  \BibitemOpen
  \bibfield  {author} {\bibinfo {author} {\bibfnamefont {T.~E.}\ \bibnamefont
  {Weller}}, \bibinfo {author} {\bibfnamefont {M.}~\bibnamefont {Ellerby}},
  \bibinfo {author} {\bibfnamefont {S.~S.}\ \bibnamefont {Saxena}}, \bibinfo
  {author} {\bibfnamefont {R.~P.}\ \bibnamefont {Smith}}, \ and\ \bibinfo
  {author} {\bibfnamefont {N.~T.}\ \bibnamefont {Skipper}},\ }\href {\doibase
  10.1038/nphys0010} {\bibfield  {journal} {\bibinfo  {journal} {Nat. Phys.}\
  }\textbf {\bibinfo {volume} {1}},\ \bibinfo {pages} {39} (\bibinfo {year}
  {2005})}\BibitemShut {NoStop}%
\bibitem [{\citenamefont {Iavarone}\ \emph {et~al.}(2003)\citenamefont
  {Iavarone}, \citenamefont {Karapetrov}, \citenamefont {Koshelev},
  \citenamefont {Kwok}, \citenamefont {Hinks}, \citenamefont {Crabtree},
  \citenamefont {Kang}, \citenamefont {Choi}, \citenamefont {Kim},\ and\
  \citenamefont {Lee}}]{0953-2048-16-2-305}%
  \BibitemOpen
  \bibfield  {author} {\bibinfo {author} {\bibfnamefont {M.}~\bibnamefont
  {Iavarone}}, \bibinfo {author} {\bibfnamefont {G.}~\bibnamefont
  {Karapetrov}}, \bibinfo {author} {\bibfnamefont {A.}~\bibnamefont
  {Koshelev}}, \bibinfo {author} {\bibfnamefont {W.~K.}\ \bibnamefont {Kwok}},
  \bibinfo {author} {\bibfnamefont {D.}~\bibnamefont {Hinks}}, \bibinfo
  {author} {\bibfnamefont {G.~W.}\ \bibnamefont {Crabtree}}, \bibinfo {author}
  {\bibfnamefont {W.~N.}\ \bibnamefont {Kang}}, \bibinfo {author}
  {\bibfnamefont {E.-M.}\ \bibnamefont {Choi}}, \bibinfo {author}
  {\bibfnamefont {H.~J.}\ \bibnamefont {Kim}}, \ and\ \bibinfo {author}
  {\bibfnamefont {S.-I.}\ \bibnamefont {Lee}},\ }\href
  {http://stacks.iop.org/0953-2048/16/i=2/a=305} {\bibfield  {journal}
  {\bibinfo  {journal} {Superconductor Science and Technology}\ }\textbf
  {\bibinfo {volume} {16}},\ \bibinfo {pages} {156} (\bibinfo {year}
  {2003})}\BibitemShut {NoStop}%
\bibitem [{\citenamefont {Choi}\ \emph {et~al.}(2002)\citenamefont {Choi},
  \citenamefont {Roundy}, \citenamefont {Sun}, \citenamefont {Cohen},\ and\
  \citenamefont {Louie}}]{Choi2002}%
  \BibitemOpen
  \bibfield  {author} {\bibinfo {author} {\bibfnamefont {H.~J.}\ \bibnamefont
  {Choi}}, \bibinfo {author} {\bibfnamefont {D.}~\bibnamefont {Roundy}},
  \bibinfo {author} {\bibfnamefont {H.}~\bibnamefont {Sun}}, \bibinfo {author}
  {\bibfnamefont {M.~L.}\ \bibnamefont {Cohen}}, \ and\ \bibinfo {author}
  {\bibfnamefont {S.~G.}\ \bibnamefont {Louie}},\ }\href {\doibase
  10.1038/nature00898} {\bibfield  {journal} {\bibinfo  {journal} {Nature}\
  }\textbf {\bibinfo {volume} {418}},\ \bibinfo {pages} {758} (\bibinfo {year}
  {2002})}\BibitemShut {NoStop}%
\bibitem [{\citenamefont {Mou}\ \emph {et~al.}(2015)\citenamefont {Mou},
  \citenamefont {Jiang}, \citenamefont {Taufour}, \citenamefont {Bud'ko},
  \citenamefont {Canfield},\ and\ \citenamefont
  {Kaminski}}]{PhysRevB.91.214519}%
  \BibitemOpen
  \bibfield  {author} {\bibinfo {author} {\bibfnamefont {D.}~\bibnamefont
  {Mou}}, \bibinfo {author} {\bibfnamefont {R.}~\bibnamefont {Jiang}}, \bibinfo
  {author} {\bibfnamefont {V.}~\bibnamefont {Taufour}}, \bibinfo {author}
  {\bibfnamefont {S.~L.}\ \bibnamefont {Bud'ko}}, \bibinfo {author}
  {\bibfnamefont {P.~C.}\ \bibnamefont {Canfield}}, \ and\ \bibinfo {author}
  {\bibfnamefont {A.}~\bibnamefont {Kaminski}},\ }\href {\doibase
  10.1103/PhysRevB.91.214519} {\bibfield  {journal} {\bibinfo  {journal} {Phys.
  Rev. B}\ }\textbf {\bibinfo {volume} {91}},\ \bibinfo {pages} {214519}
  (\bibinfo {year} {2015})}\BibitemShut {NoStop}%
\bibitem [{\citenamefont {Margine}\ and\ \citenamefont
  {Giustino}(2013)}]{PhysRevB.87.024505}%
  \BibitemOpen
  \bibfield  {author} {\bibinfo {author} {\bibfnamefont {E.~R.}\ \bibnamefont
  {Margine}}\ and\ \bibinfo {author} {\bibfnamefont {F.}~\bibnamefont
  {Giustino}},\ }\href {\doibase 10.1103/PhysRevB.87.024505} {\bibfield
  {journal} {\bibinfo  {journal} {Phys. Rev. B}\ }\textbf {\bibinfo {volume}
  {87}},\ \bibinfo {pages} {024505} (\bibinfo {year} {2013})}\BibitemShut
  {NoStop}%
\bibitem [{\citenamefont {Aperis}\ \emph {et~al.}(2015)\citenamefont {Aperis},
  \citenamefont {Maldonado},\ and\ \citenamefont
  {Oppeneer}}]{PhysRevB.92.054516}%
  \BibitemOpen
  \bibfield  {author} {\bibinfo {author} {\bibfnamefont {A.}~\bibnamefont
  {Aperis}}, \bibinfo {author} {\bibfnamefont {P.}~\bibnamefont {Maldonado}}, \
  and\ \bibinfo {author} {\bibfnamefont {P.~M.}\ \bibnamefont {Oppeneer}},\
  }\href {\doibase 10.1103/PhysRevB.92.054516} {\bibfield  {journal} {\bibinfo
  {journal} {Phys. Rev. B}\ }\textbf {\bibinfo {volume} {92}},\ \bibinfo
  {pages} {054516} (\bibinfo {year} {2015})}\BibitemShut {NoStop}%
\bibitem [{\citenamefont {Milo\v{s}evi\'{c}}\ and\ \citenamefont
  {Perali}(2015)}]{0953-2048-28-6-060201}%
  \BibitemOpen
  \bibfield  {author} {\bibinfo {author} {\bibfnamefont {M.~V.}\ \bibnamefont
  {Milo\v{s}evi\'{c}}}\ and\ \bibinfo {author} {\bibfnamefont {A.}~\bibnamefont
  {Perali}},\ }\href {http://stacks.iop.org/0953-2048/28/i=6/a=060201}
  {\bibfield  {journal} {\bibinfo  {journal} {Supercond. Sci. Technol.}\
  }\textbf {\bibinfo {volume} {28}},\ \bibinfo {pages} {060201} (\bibinfo
  {year} {2015})}\BibitemShut {NoStop}%
\bibitem [{\citenamefont {Stanev}\ and\ \citenamefont {Te\ifmmode
  \check{s}\else \v{s}\fi{}anovi\ifmmode~\acute{c}\else
  \'{c}\fi{}}(2010)}]{PhysRevB.81.134522}%
  \BibitemOpen
  \bibfield  {author} {\bibinfo {author} {\bibfnamefont {V.}~\bibnamefont
  {Stanev}}\ and\ \bibinfo {author} {\bibfnamefont {Z.}~\bibnamefont
  {Te\ifmmode \check{s}\else \v{s}\fi{}anovi\ifmmode~\acute{c}\else
  \'{c}\fi{}}},\ }\href {\doibase 10.1103/PhysRevB.81.134522} {\bibfield
  {journal} {\bibinfo  {journal} {Phys. Rev. B}\ }\textbf {\bibinfo {volume}
  {81}},\ \bibinfo {pages} {134522} (\bibinfo {year} {2010})}\BibitemShut
  {NoStop}%
\bibitem [{\citenamefont {Babaev}(2002)}]{PhysRevLett.89.067001}%
  \BibitemOpen
  \bibfield  {author} {\bibinfo {author} {\bibfnamefont {E.}~\bibnamefont
  {Babaev}},\ }\href {\doibase 10.1103/PhysRevLett.89.067001} {\bibfield
  {journal} {\bibinfo  {journal} {Phys. Rev. Lett.}\ }\textbf {\bibinfo
  {volume} {89}},\ \bibinfo {pages} {067001} (\bibinfo {year}
  {2002})}\BibitemShut {NoStop}%
\bibitem [{\citenamefont {Garaud}\ \emph {et~al.}(2011)\citenamefont {Garaud},
  \citenamefont {Carlstr\"om},\ and\ \citenamefont
  {Babaev}}]{PhysRevLett.107.197001}%
  \BibitemOpen
  \bibfield  {author} {\bibinfo {author} {\bibfnamefont {J.}~\bibnamefont
  {Garaud}}, \bibinfo {author} {\bibfnamefont {J.}~\bibnamefont {Carlstr\"om}},
  \ and\ \bibinfo {author} {\bibfnamefont {E.}~\bibnamefont {Babaev}},\ }\href
  {\doibase 10.1103/PhysRevLett.107.197001} {\bibfield  {journal} {\bibinfo
  {journal} {Phys. Rev. Lett.}\ }\textbf {\bibinfo {volume} {107}},\ \bibinfo
  {pages} {197001} (\bibinfo {year} {2011})}\BibitemShut {NoStop}%
\bibitem [{\citenamefont {da~Silva}\ \emph {et~al.}(2015)\citenamefont
  {da~Silva}, \citenamefont {Milo\v{s}evi\'{c}}, \citenamefont {Shanenko},
  \citenamefont {Peeters},\ and\ \citenamefont {Aguiar}}]{rogerio}%
  \BibitemOpen
  \bibfield  {author} {\bibinfo {author} {\bibfnamefont {R.~M.}\ \bibnamefont
  {da~Silva}}, \bibinfo {author} {\bibfnamefont {M.~V.}\ \bibnamefont
  {Milo\v{s}evi\'{c}}}, \bibinfo {author} {\bibfnamefont {A.~A.}\ \bibnamefont
  {Shanenko}}, \bibinfo {author} {\bibfnamefont {F.~M.}\ \bibnamefont
  {Peeters}}, \ and\ \bibinfo {author} {\bibfnamefont {J.~A.}\ \bibnamefont
  {Aguiar}},\ }\href@noop {} {\bibfield  {journal} {\bibinfo  {journal} {Sci.
  Rep.}\ }\textbf {\bibinfo {volume} {5}},\ \bibinfo {pages} {12695} (\bibinfo
  {year} {2015})}\BibitemShut {NoStop}%
\bibitem [{\citenamefont {Komendov\'a}\ \emph {et~al.}(2012)\citenamefont
  {Komendov\'a}, \citenamefont {Chen}, \citenamefont {Shanenko}, \citenamefont
  {Milo\ifmmode \check{s}\else \v{s}\fi{}evi\ifmmode~\acute{c}\else
  \'{c}\fi{}},\ and\ \citenamefont {Peeters}}]{PhysRevLett.108.207002}%
  \BibitemOpen
  \bibfield  {author} {\bibinfo {author} {\bibfnamefont {L.}~\bibnamefont
  {Komendov\'a}}, \bibinfo {author} {\bibfnamefont {Y.}~\bibnamefont {Chen}},
  \bibinfo {author} {\bibfnamefont {A.~A.}\ \bibnamefont {Shanenko}}, \bibinfo
  {author} {\bibfnamefont {M.~V.}\ \bibnamefont {Milo\ifmmode \check{s}\else
  \v{s}\fi{}evi\ifmmode~\acute{c}\else \'{c}\fi{}}}, \ and\ \bibinfo {author}
  {\bibfnamefont {F.~M.}\ \bibnamefont {Peeters}},\ }\href {\doibase
  10.1103/PhysRevLett.108.207002} {\bibfield  {journal} {\bibinfo  {journal}
  {Phys. Rev. Lett.}\ }\textbf {\bibinfo {volume} {108}},\ \bibinfo {pages}
  {207002} (\bibinfo {year} {2012})}\BibitemShut {NoStop}%
\bibitem [{\citenamefont {Orlova}\ \emph {et~al.}(2013)\citenamefont {Orlova},
  \citenamefont {Shanenko}, \citenamefont {Milo\ifmmode \check{s}\else
  \v{s}\fi{}evi\ifmmode~\acute{c}\else \'{c}\fi{}}, \citenamefont {Peeters},
  \citenamefont {Vagov},\ and\ \citenamefont {Axt}}]{PhysRevB.87.134510}%
  \BibitemOpen
  \bibfield  {author} {\bibinfo {author} {\bibfnamefont {N.~V.}\ \bibnamefont
  {Orlova}}, \bibinfo {author} {\bibfnamefont {A.~A.}\ \bibnamefont
  {Shanenko}}, \bibinfo {author} {\bibfnamefont {M.~V.}\ \bibnamefont
  {Milo\ifmmode \check{s}\else \v{s}\fi{}evi\ifmmode~\acute{c}\else
  \'{c}\fi{}}}, \bibinfo {author} {\bibfnamefont {F.~M.}\ \bibnamefont
  {Peeters}}, \bibinfo {author} {\bibfnamefont {A.~V.}\ \bibnamefont {Vagov}},
  \ and\ \bibinfo {author} {\bibfnamefont {V.~M.}\ \bibnamefont {Axt}},\ }\href
  {\doibase 10.1103/PhysRevB.87.134510} {\bibfield  {journal} {\bibinfo
  {journal} {Phys. Rev. B}\ }\textbf {\bibinfo {volume} {87}},\ \bibinfo
  {pages} {134510} (\bibinfo {year} {2013})}\BibitemShut {NoStop}%
\bibitem [{\citenamefont {Linscheid}\ \emph {et~al.}(2015)\citenamefont
  {Linscheid}, \citenamefont {Sanna}, \citenamefont {Floris},\ and\
  \citenamefont {Gross}}]{PhysRevLett.115.097002}%
  \BibitemOpen
  \bibfield  {author} {\bibinfo {author} {\bibfnamefont {A.}~\bibnamefont
  {Linscheid}}, \bibinfo {author} {\bibfnamefont {A.}~\bibnamefont {Sanna}},
  \bibinfo {author} {\bibfnamefont {A.}~\bibnamefont {Floris}}, \ and\ \bibinfo
  {author} {\bibfnamefont {E.~K.~U.}\ \bibnamefont {Gross}},\ }\href {\doibase
  10.1103/PhysRevLett.115.097002} {\bibfield  {journal} {\bibinfo  {journal}
  {Phys. Rev. Lett.}\ }\textbf {\bibinfo {volume} {115}},\ \bibinfo {pages}
  {097002} (\bibinfo {year} {2015})}\BibitemShut {NoStop}%
\bibitem [{\citenamefont {Cudazzo}\ \emph {et~al.}(2008)\citenamefont
  {Cudazzo}, \citenamefont {Profeta}, \citenamefont {Sanna}, \citenamefont
  {Floris}, \citenamefont {Continenza}, \citenamefont {Massidda},\ and\
  \citenamefont {Gross}}]{PhysRevLett.100.257001}%
  \BibitemOpen
  \bibfield  {author} {\bibinfo {author} {\bibfnamefont {P.}~\bibnamefont
  {Cudazzo}}, \bibinfo {author} {\bibfnamefont {G.}~\bibnamefont {Profeta}},
  \bibinfo {author} {\bibfnamefont {A.}~\bibnamefont {Sanna}}, \bibinfo
  {author} {\bibfnamefont {A.}~\bibnamefont {Floris}}, \bibinfo {author}
  {\bibfnamefont {A.}~\bibnamefont {Continenza}}, \bibinfo {author}
  {\bibfnamefont {S.}~\bibnamefont {Massidda}}, \ and\ \bibinfo {author}
  {\bibfnamefont {E.~K.~U.}\ \bibnamefont {Gross}},\ }\href {\doibase
  10.1103/PhysRevLett.100.257001} {\bibfield  {journal} {\bibinfo  {journal}
  {Phys. Rev. Lett.}\ }\textbf {\bibinfo {volume} {100}},\ \bibinfo {pages}
  {257001} (\bibinfo {year} {2008})}\BibitemShut {NoStop}%
\bibitem [{\citenamefont {Bersier}\ \emph {et~al.}(2009)\citenamefont
  {Bersier}, \citenamefont {Floris}, \citenamefont {Sanna}, \citenamefont
  {Profeta}, \citenamefont {Continenza}, \citenamefont {Gross},\ and\
  \citenamefont {Massidda}}]{PhysRevB.79.104503}%
  \BibitemOpen
  \bibfield  {author} {\bibinfo {author} {\bibfnamefont {C.}~\bibnamefont
  {Bersier}}, \bibinfo {author} {\bibfnamefont {A.}~\bibnamefont {Floris}},
  \bibinfo {author} {\bibfnamefont {A.}~\bibnamefont {Sanna}}, \bibinfo
  {author} {\bibfnamefont {G.}~\bibnamefont {Profeta}}, \bibinfo {author}
  {\bibfnamefont {A.}~\bibnamefont {Continenza}}, \bibinfo {author}
  {\bibfnamefont {E.~K.~U.}\ \bibnamefont {Gross}}, \ and\ \bibinfo {author}
  {\bibfnamefont {S.}~\bibnamefont {Massidda}},\ }\href {\doibase
  10.1103/PhysRevB.79.104503} {\bibfield  {journal} {\bibinfo  {journal} {Phys.
  Rev. B}\ }\textbf {\bibinfo {volume} {79}},\ \bibinfo {pages} {104503}
  (\bibinfo {year} {2009})}\BibitemShut {NoStop}%
\bibitem [{\citenamefont {Brun}\ \emph {et~al.}(2017)\citenamefont {Brun},
  \citenamefont {Cren},\ and\ \citenamefont
  {Roditchev}}]{0953-2048-30-1-013003}%
  \BibitemOpen
  \bibfield  {author} {\bibinfo {author} {\bibfnamefont {C.}~\bibnamefont
  {Brun}}, \bibinfo {author} {\bibfnamefont {T.}~\bibnamefont {Cren}}, \ and\
  \bibinfo {author} {\bibfnamefont {D.}~\bibnamefont {Roditchev}},\ }\href
  {http://stacks.iop.org/0953-2048/30/i=1/a=013003} {\bibfield  {journal}
  {\bibinfo  {journal} {Supercond. Sci. Technol.}\ }\textbf {\bibinfo {volume}
  {30}},\ \bibinfo {pages} {013003} (\bibinfo {year} {2017})}\BibitemShut
  {NoStop}%
\bibitem [{\citenamefont {Uchihashi}(2017)}]{0953-2048-30-1-013002}%
  \BibitemOpen
  \bibfield  {author} {\bibinfo {author} {\bibfnamefont {T.}~\bibnamefont
  {Uchihashi}},\ }\href {http://stacks.iop.org/0953-2048/30/i=1/a=013002}
  {\bibfield  {journal} {\bibinfo  {journal} {Supercond. Sci. Technol.}\
  }\textbf {\bibinfo {volume} {30}},\ \bibinfo {pages} {013002} (\bibinfo
  {year} {2017})}\BibitemShut {NoStop}%
\bibitem [{\citenamefont {Qin}\ \emph {et~al.}(2009)\citenamefont {Qin},
  \citenamefont {Kim}, \citenamefont {Niu},\ and\ \citenamefont
  {Shih}}]{Qin2009}%
  \BibitemOpen
  \bibfield  {author} {\bibinfo {author} {\bibfnamefont {S.}~\bibnamefont
  {Qin}}, \bibinfo {author} {\bibfnamefont {J.}~\bibnamefont {Kim}}, \bibinfo
  {author} {\bibfnamefont {Q.}~\bibnamefont {Niu}}, \ and\ \bibinfo {author}
  {\bibfnamefont {C.}~\bibnamefont {Shih}},\ }\href@noop {} {\bibfield
  {journal} {\bibinfo  {journal} {Science}\ }\textbf {\bibinfo {volume}
  {324}},\ \bibinfo {pages} {1314} (\bibinfo {year} {2009})}\BibitemShut
  {NoStop}%
\bibitem [{\citenamefont {Zhang}\ \emph {et~al.}(2010)\citenamefont {Zhang},
  \citenamefont {Cheng}, \citenamefont {Li}, \citenamefont {Sun}, \citenamefont
  {Wang}, \citenamefont {Zhu}, \citenamefont {He}, \citenamefont {Wang},
  \citenamefont {Ma}, \citenamefont {Chen}, \citenamefont {Wang}, \citenamefont
  {Liu}, \citenamefont {Lin}, \citenamefont {Jia},\ and\ \citenamefont
  {Xue}}]{Zhang2010}%
  \BibitemOpen
  \bibfield  {author} {\bibinfo {author} {\bibfnamefont {T.}~\bibnamefont
  {Zhang}}, \bibinfo {author} {\bibfnamefont {P.}~\bibnamefont {Cheng}},
  \bibinfo {author} {\bibfnamefont {W.}~\bibnamefont {Li}}, \bibinfo {author}
  {\bibfnamefont {Y.}~\bibnamefont {Sun}}, \bibinfo {author} {\bibfnamefont
  {G.}~\bibnamefont {Wang}}, \bibinfo {author} {\bibfnamefont {X.}~\bibnamefont
  {Zhu}}, \bibinfo {author} {\bibfnamefont {K.}~\bibnamefont {He}}, \bibinfo
  {author} {\bibfnamefont {L.}~\bibnamefont {Wang}}, \bibinfo {author}
  {\bibfnamefont {X.}~\bibnamefont {Ma}}, \bibinfo {author} {\bibfnamefont
  {X.}~\bibnamefont {Chen}}, \bibinfo {author} {\bibfnamefont {Y.}~\bibnamefont
  {Wang}}, \bibinfo {author} {\bibfnamefont {Y.}~\bibnamefont {Liu}}, \bibinfo
  {author} {\bibfnamefont {H.}~\bibnamefont {Lin}}, \bibinfo {author}
  {\bibfnamefont {J.}~\bibnamefont {Jia}}, \ and\ \bibinfo {author}
  {\bibfnamefont {Q.}~\bibnamefont {Xue}},\ }\href {\doibase 10.1038/nphys1499}
  {\bibfield  {journal} {\bibinfo  {journal} {Nat. Phys.}\ }\textbf {\bibinfo
  {volume} {6}},\ \bibinfo {pages} {104} (\bibinfo {year} {2010})}\BibitemShut
  {NoStop}%
\bibitem [{\citenamefont {Cao}\ \emph {et~al.}(2015)\citenamefont {Cao},
  \citenamefont {Mishchenko}, \citenamefont {Yu}, \citenamefont {Khestanova},
  \citenamefont {Rooney}, \citenamefont {Prestat}, \citenamefont {Kretinin},
  \citenamefont {Blake}, \citenamefont {Shalom}, \citenamefont {Woods},
  \citenamefont {Chapman}, \citenamefont {Balakrishnan}, \citenamefont
  {Grigorieva}, \citenamefont {Novoselov}, \citenamefont {Piot}, \citenamefont
  {Potemski}, \citenamefont {Watanabe}, \citenamefont {Taniguchi},
  \citenamefont {Haigh}, \citenamefont {Geim},\ and\ \citenamefont
  {Gorbachev}}]{doi:10.1021/acs.nanolett.5b00648}%
  \BibitemOpen
  \bibfield  {author} {\bibinfo {author} {\bibfnamefont {Y.}~\bibnamefont
  {Cao}}, \bibinfo {author} {\bibfnamefont {A.}~\bibnamefont {Mishchenko}},
  \bibinfo {author} {\bibfnamefont {G.~L.}\ \bibnamefont {Yu}}, \bibinfo
  {author} {\bibfnamefont {E.}~\bibnamefont {Khestanova}}, \bibinfo {author}
  {\bibfnamefont {A.~P.}\ \bibnamefont {Rooney}}, \bibinfo {author}
  {\bibfnamefont {E.}~\bibnamefont {Prestat}}, \bibinfo {author} {\bibfnamefont
  {A.~V.}\ \bibnamefont {Kretinin}}, \bibinfo {author} {\bibfnamefont
  {P.}~\bibnamefont {Blake}}, \bibinfo {author} {\bibfnamefont {M.~B.}\
  \bibnamefont {Shalom}}, \bibinfo {author} {\bibfnamefont {C.}~\bibnamefont
  {Woods}}, \bibinfo {author} {\bibfnamefont {J.}~\bibnamefont {Chapman}},
  \bibinfo {author} {\bibfnamefont {G.}~\bibnamefont {Balakrishnan}}, \bibinfo
  {author} {\bibfnamefont {I.~V.}\ \bibnamefont {Grigorieva}}, \bibinfo
  {author} {\bibfnamefont {K.~S.}\ \bibnamefont {Novoselov}}, \bibinfo {author}
  {\bibfnamefont {B.~A.}\ \bibnamefont {Piot}}, \bibinfo {author}
  {\bibfnamefont {M.}~\bibnamefont {Potemski}}, \bibinfo {author}
  {\bibfnamefont {K.}~\bibnamefont {Watanabe}}, \bibinfo {author}
  {\bibfnamefont {T.}~\bibnamefont {Taniguchi}}, \bibinfo {author}
  {\bibfnamefont {S.~J.}\ \bibnamefont {Haigh}}, \bibinfo {author}
  {\bibfnamefont {A.~K.}\ \bibnamefont {Geim}}, \ and\ \bibinfo {author}
  {\bibfnamefont {R.~V.}\ \bibnamefont {Gorbachev}},\ }\href {\doibase
  10.1021/acs.nanolett.5b00648} {\bibfield  {journal} {\bibinfo  {journal}
  {Nano Lett.}\ }\textbf {\bibinfo {volume} {15}},\ \bibinfo {pages} {4914}
  (\bibinfo {year} {2015})}\BibitemShut {NoStop}%
\bibitem [{\citenamefont {Ugeda}\ \emph {et~al.}(2016)\citenamefont {Ugeda},
  \citenamefont {Bradley}, \citenamefont {Zhang}, \citenamefont {Onishi},
  \citenamefont {Chen}, \citenamefont {Ruan}, \citenamefont
  {Ojeda-Aristizabal}, \citenamefont {Ryu}, \citenamefont {Edmonds},
  \citenamefont {Tsai}, \citenamefont {Riss}, \citenamefont {Mo}, \citenamefont
  {Lee}, \citenamefont {Zettl}, \citenamefont {Hussain}, \citenamefont {Shen},\
  and\ \citenamefont {Crommie}}]{Ugeda2016}%
  \BibitemOpen
  \bibfield  {author} {\bibinfo {author} {\bibfnamefont {M.~M.}\ \bibnamefont
  {Ugeda}}, \bibinfo {author} {\bibfnamefont {A.~J.}\ \bibnamefont {Bradley}},
  \bibinfo {author} {\bibfnamefont {Y.}~\bibnamefont {Zhang}}, \bibinfo
  {author} {\bibfnamefont {S.}~\bibnamefont {Onishi}}, \bibinfo {author}
  {\bibfnamefont {Y.}~\bibnamefont {Chen}}, \bibinfo {author} {\bibfnamefont
  {W.}~\bibnamefont {Ruan}}, \bibinfo {author} {\bibfnamefont {C.}~\bibnamefont
  {Ojeda-Aristizabal}}, \bibinfo {author} {\bibfnamefont {H.}~\bibnamefont
  {Ryu}}, \bibinfo {author} {\bibfnamefont {M.~T.}\ \bibnamefont {Edmonds}},
  \bibinfo {author} {\bibfnamefont {H.-Z.}\ \bibnamefont {Tsai}}, \bibinfo
  {author} {\bibfnamefont {A.}~\bibnamefont {Riss}}, \bibinfo {author}
  {\bibfnamefont {S.-K.}\ \bibnamefont {Mo}}, \bibinfo {author} {\bibfnamefont
  {D.}~\bibnamefont {Lee}}, \bibinfo {author} {\bibfnamefont {A.}~\bibnamefont
  {Zettl}}, \bibinfo {author} {\bibfnamefont {Z.}~\bibnamefont {Hussain}},
  \bibinfo {author} {\bibfnamefont {Z.-X.}\ \bibnamefont {Shen}}, \ and\
  \bibinfo {author} {\bibfnamefont {M.~F.}\ \bibnamefont {Crommie}},\ }\href
  {http://dx.doi.org/10.1038/nphys3527} {\bibfield  {journal} {\bibinfo
  {journal} {Nat. Phys.}\ }\textbf {\bibinfo {volume} {12}},\ \bibinfo {pages}
  {92} (\bibinfo {year} {2016})}\BibitemShut {NoStop}%
\bibitem [{\citenamefont {Xi}\ \emph {et~al.}(2016)\citenamefont {Xi},
  \citenamefont {Wang}, \citenamefont {Zhao}, \citenamefont {Park},
  \citenamefont {Law}, \citenamefont {Berger}, \citenamefont {Forro},
  \citenamefont {Shan},\ and\ \citenamefont {Mak}}]{Xi2016}%
  \BibitemOpen
  \bibfield  {author} {\bibinfo {author} {\bibfnamefont {X.}~\bibnamefont
  {Xi}}, \bibinfo {author} {\bibfnamefont {Z.}~\bibnamefont {Wang}}, \bibinfo
  {author} {\bibfnamefont {W.}~\bibnamefont {Zhao}}, \bibinfo {author}
  {\bibfnamefont {J.-H.}\ \bibnamefont {Park}}, \bibinfo {author}
  {\bibfnamefont {K.~T.}\ \bibnamefont {Law}}, \bibinfo {author} {\bibfnamefont
  {H.}~\bibnamefont {Berger}}, \bibinfo {author} {\bibfnamefont
  {L.}~\bibnamefont {Forro}}, \bibinfo {author} {\bibfnamefont
  {J.}~\bibnamefont {Shan}}, \ and\ \bibinfo {author} {\bibfnamefont {K.~F.}\
  \bibnamefont {Mak}},\ }\href {http://dx.doi.org/10.1038/nphys3538} {\bibfield
   {journal} {\bibinfo  {journal} {Nat. Phys.}\ }\textbf {\bibinfo {volume}
  {12}},\ \bibinfo {pages} {139} (\bibinfo {year} {2016})}\BibitemShut
  {NoStop}%
\bibitem [{\citenamefont {Profeta}\ \emph {et~al.}(2012)\citenamefont
  {Profeta}, \citenamefont {Calandra},\ and\ \citenamefont
  {Mauri}}]{Profeta2012}%
  \BibitemOpen
  \bibfield  {author} {\bibinfo {author} {\bibfnamefont {G.}~\bibnamefont
  {Profeta}}, \bibinfo {author} {\bibfnamefont {M.}~\bibnamefont {Calandra}}, \
  and\ \bibinfo {author} {\bibfnamefont {F.}~\bibnamefont {Mauri}},\ }\href
  {\doibase 10.1038/nphys2181} {\bibfield  {journal} {\bibinfo  {journal} {Nat.
  Phys.}\ }\textbf {\bibinfo {volume} {8}},\ \bibinfo {pages} {131} (\bibinfo
  {year} {2012})}\BibitemShut {NoStop}%
\bibitem [{\citenamefont {Guzman}\ \emph {et~al.}(2014)\citenamefont {Guzman},
  \citenamefont {Alyahyaei},\ and\ \citenamefont
  {Jishi}}]{2053-1583-1-2-021005}%
  \BibitemOpen
  \bibfield  {author} {\bibinfo {author} {\bibfnamefont {D.~M.}\ \bibnamefont
  {Guzman}}, \bibinfo {author} {\bibfnamefont {H.~M.}\ \bibnamefont
  {Alyahyaei}}, \ and\ \bibinfo {author} {\bibfnamefont {R.~A.}\ \bibnamefont
  {Jishi}},\ }\href {http://stacks.iop.org/2053-1583/1/i=2/a=021005} {\bibfield
   {journal} {\bibinfo  {journal} {2D Mater.}\ }\textbf {\bibinfo {volume}
  {1}},\ \bibinfo {pages} {021005} (\bibinfo {year} {2014})}\BibitemShut
  {NoStop}%
\bibitem [{\citenamefont {Ludbrook}\ \emph {et~al.}(2015)\citenamefont
  {Ludbrook}, \citenamefont {Levy}, \citenamefont {Nigge}, \citenamefont
  {Zonno}, \citenamefont {Schneider}, \citenamefont {Dvorak}, \citenamefont
  {Veenstra}, \citenamefont {Zhdanovich}, \citenamefont {Wong}, \citenamefont
  {Dosanjh}, \citenamefont {Straßer}, \citenamefont {Stöhr}, \citenamefont
  {Forti}, \citenamefont {Ast}, \citenamefont {Starke},\ and\ \citenamefont
  {Damascelli}}]{Ludbrook22092015}%
  \BibitemOpen
  \bibfield  {author} {\bibinfo {author} {\bibfnamefont {B.~M.}\ \bibnamefont
  {Ludbrook}}, \bibinfo {author} {\bibfnamefont {G.}~\bibnamefont {Levy}},
  \bibinfo {author} {\bibfnamefont {P.}~\bibnamefont {Nigge}}, \bibinfo
  {author} {\bibfnamefont {M.}~\bibnamefont {Zonno}}, \bibinfo {author}
  {\bibfnamefont {M.}~\bibnamefont {Schneider}}, \bibinfo {author}
  {\bibfnamefont {D.~J.}\ \bibnamefont {Dvorak}}, \bibinfo {author}
  {\bibfnamefont {C.~N.}\ \bibnamefont {Veenstra}}, \bibinfo {author}
  {\bibfnamefont {S.}~\bibnamefont {Zhdanovich}}, \bibinfo {author}
  {\bibfnamefont {D.}~\bibnamefont {Wong}}, \bibinfo {author} {\bibfnamefont
  {P.}~\bibnamefont {Dosanjh}}, \bibinfo {author} {\bibfnamefont
  {C.}~\bibnamefont {Straßer}}, \bibinfo {author} {\bibfnamefont
  {A.}~\bibnamefont {Stöhr}}, \bibinfo {author} {\bibfnamefont
  {S.}~\bibnamefont {Forti}}, \bibinfo {author} {\bibfnamefont {C.~R.}\
  \bibnamefont {Ast}}, \bibinfo {author} {\bibfnamefont {U.}~\bibnamefont
  {Starke}}, \ and\ \bibinfo {author} {\bibfnamefont {A.}~\bibnamefont
  {Damascelli}},\ }\href {\doibase 10.1073/pnas.1510435112} {\bibfield
  {journal} {\bibinfo  {journal} {Proc. Natl. Acad. Sci.}\ }\textbf {\bibinfo
  {volume} {112}},\ \bibinfo {pages} {11795} (\bibinfo {year}
  {2015})}\BibitemShut {NoStop}%
\bibitem [{\citenamefont {Kanetani}\ \emph {et~al.}(2012)\citenamefont
  {Kanetani}, \citenamefont {Sugawara}, \citenamefont {Sato}, \citenamefont
  {Shimizu}, \citenamefont {Iwaya}, \citenamefont {Hitosugi},\ and\
  \citenamefont {Takahashi}}]{Kanetani27112012}%
  \BibitemOpen
  \bibfield  {author} {\bibinfo {author} {\bibfnamefont {K.}~\bibnamefont
  {Kanetani}}, \bibinfo {author} {\bibfnamefont {K.}~\bibnamefont {Sugawara}},
  \bibinfo {author} {\bibfnamefont {T.}~\bibnamefont {Sato}}, \bibinfo {author}
  {\bibfnamefont {R.}~\bibnamefont {Shimizu}}, \bibinfo {author} {\bibfnamefont
  {K.}~\bibnamefont {Iwaya}}, \bibinfo {author} {\bibfnamefont
  {T.}~\bibnamefont {Hitosugi}}, \ and\ \bibinfo {author} {\bibfnamefont
  {T.}~\bibnamefont {Takahashi}},\ }\href {\doibase 10.1073/pnas.1208889109}
  {\bibfield  {journal} {\bibinfo  {journal} {Proc. Natl. Acad. Sci.}\ }\textbf
  {\bibinfo {volume} {109}},\ \bibinfo {pages} {19610} (\bibinfo {year}
  {2012})}\BibitemShut {NoStop}%
\bibitem [{\citenamefont {Chapman}\ \emph {et~al.}(2016)\citenamefont
  {Chapman}, \citenamefont {Su}, \citenamefont {Howard}, \citenamefont
  {Kundys}, \citenamefont {Grigorenko}, \citenamefont {Guinea}, \citenamefont
  {Geim}, \citenamefont {Grigorieva},\ and\ \citenamefont
  {Nair}}]{Chapman2016}%
  \BibitemOpen
  \bibfield  {author} {\bibinfo {author} {\bibfnamefont {J.}~\bibnamefont
  {Chapman}}, \bibinfo {author} {\bibfnamefont {Y.}~\bibnamefont {Su}},
  \bibinfo {author} {\bibfnamefont {C.~A.}\ \bibnamefont {Howard}}, \bibinfo
  {author} {\bibfnamefont {D.}~\bibnamefont {Kundys}}, \bibinfo {author}
  {\bibfnamefont {A.~N.}\ \bibnamefont {Grigorenko}}, \bibinfo {author}
  {\bibfnamefont {F.}~\bibnamefont {Guinea}}, \bibinfo {author} {\bibfnamefont
  {A.~K.}\ \bibnamefont {Geim}}, \bibinfo {author} {\bibfnamefont {I.~V.}\
  \bibnamefont {Grigorieva}}, \ and\ \bibinfo {author} {\bibfnamefont {R.~R.}\
  \bibnamefont {Nair}},\ }\href {http://dx.doi.org/10.1038/srep23254}
  {\bibfield  {journal} {\bibinfo  {journal} {Sci. Rep.}\ }\textbf {\bibinfo
  {volume} {6}},\ \bibinfo {pages} {23254} (\bibinfo {year}
  {2016})}\BibitemShut {NoStop}%
\bibitem [{\citenamefont {Verbitskiy}\ \emph {et~al.}(2016)\citenamefont
  {Verbitskiy}, \citenamefont {Fedorov}, \citenamefont {Tresca}, \citenamefont
  {Profeta}, \citenamefont {Petaccia}, \citenamefont {Senkovskiy},
  \citenamefont {Usachov}, \citenamefont {Vyalikh}, \citenamefont {Yashina},
  \citenamefont {Eliseev}, \citenamefont {Pichler},\ and\ \citenamefont
  {Grüneis}}]{2053-1583-3-4-045003}%
  \BibitemOpen
  \bibfield  {author} {\bibinfo {author} {\bibfnamefont {N.~I.}\ \bibnamefont
  {Verbitskiy}}, \bibinfo {author} {\bibfnamefont {A.~V.}\ \bibnamefont
  {Fedorov}}, \bibinfo {author} {\bibfnamefont {C.}~\bibnamefont {Tresca}},
  \bibinfo {author} {\bibfnamefont {G.}~\bibnamefont {Profeta}}, \bibinfo
  {author} {\bibfnamefont {L.}~\bibnamefont {Petaccia}}, \bibinfo {author}
  {\bibfnamefont {B.~V.}\ \bibnamefont {Senkovskiy}}, \bibinfo {author}
  {\bibfnamefont {D.~Y.}\ \bibnamefont {Usachov}}, \bibinfo {author}
  {\bibfnamefont {D.~V.}\ \bibnamefont {Vyalikh}}, \bibinfo {author}
  {\bibfnamefont {L.~V.}\ \bibnamefont {Yashina}}, \bibinfo {author}
  {\bibfnamefont {A.~A.}\ \bibnamefont {Eliseev}}, \bibinfo {author}
  {\bibfnamefont {T.}~\bibnamefont {Pichler}}, \ and\ \bibinfo {author}
  {\bibfnamefont {A.}~\bibnamefont {Grüneis}},\ }\href
  {http://stacks.iop.org/2053-1583/3/i=4/a=045003} {\bibfield  {journal}
  {\bibinfo  {journal} {2D Mater.}\ }\textbf {\bibinfo {volume} {3}},\ \bibinfo
  {pages} {045003} (\bibinfo {year} {2016})}\BibitemShut {NoStop}%
\bibitem [{\citenamefont {Bollinger}\ \emph {et~al.}(2011)\citenamefont
  {Bollinger}, \citenamefont {Dubuis}, \citenamefont {Yoon}, \citenamefont
  {Pavuna}, \citenamefont {Misewich},\ and\ \citenamefont
  {Bozovic}}]{Bollinger2011}%
  \BibitemOpen
  \bibfield  {author} {\bibinfo {author} {\bibfnamefont {A.~T.}\ \bibnamefont
  {Bollinger}}, \bibinfo {author} {\bibfnamefont {G.}~\bibnamefont {Dubuis}},
  \bibinfo {author} {\bibfnamefont {J.}~\bibnamefont {Yoon}}, \bibinfo {author}
  {\bibfnamefont {D.}~\bibnamefont {Pavuna}}, \bibinfo {author} {\bibfnamefont
  {J.}~\bibnamefont {Misewich}}, \ and\ \bibinfo {author} {\bibfnamefont
  {I.}~\bibnamefont {Bozovic}},\ }\href {\doibase 10.1038/nature09998}
  {\bibfield  {journal} {\bibinfo  {journal} {Nature}\ }\textbf {\bibinfo
  {volume} {472}},\ \bibinfo {pages} {458} (\bibinfo {year}
  {2011})}\BibitemShut {NoStop}%
\bibitem [{\citenamefont {Ge}\ \emph {et~al.}(2015)\citenamefont {Ge},
  \citenamefont {Liu}, \citenamefont {Liu}, \citenamefont {Gao}, \citenamefont
  {Qian}, \citenamefont {Xue}, \citenamefont {Liu},\ and\ \citenamefont
  {Jia}}]{Ge2015}%
  \BibitemOpen
  \bibfield  {author} {\bibinfo {author} {\bibfnamefont {J.}~\bibnamefont
  {Ge}}, \bibinfo {author} {\bibfnamefont {Z.}~\bibnamefont {Liu}}, \bibinfo
  {author} {\bibfnamefont {C.}~\bibnamefont {Liu}}, \bibinfo {author}
  {\bibfnamefont {C.}~\bibnamefont {Gao}}, \bibinfo {author} {\bibfnamefont
  {D.}~\bibnamefont {Qian}}, \bibinfo {author} {\bibfnamefont {Q.}~\bibnamefont
  {Xue}}, \bibinfo {author} {\bibfnamefont {Y.}~\bibnamefont {Liu}}, \ and\
  \bibinfo {author} {\bibfnamefont {J.}~\bibnamefont {Jia}},\ }\href@noop {}
  {\bibfield  {journal} {\bibinfo  {journal} {Nat. Mater.}\ }\textbf {\bibinfo
  {volume} {14}},\ \bibinfo {pages} {285} (\bibinfo {year} {2015})}\BibitemShut
  {NoStop}%
\bibitem [{\citenamefont {Golod}\ \emph {et~al.}(2015)\citenamefont {Golod},
  \citenamefont {Iovan},\ and\ \citenamefont {Krasnov}}]{Golod2015}%
  \BibitemOpen
  \bibfield  {author} {\bibinfo {author} {\bibfnamefont {T.}~\bibnamefont
  {Golod}}, \bibinfo {author} {\bibfnamefont {A.}~\bibnamefont {Iovan}}, \ and\
  \bibinfo {author} {\bibfnamefont {V.~M.}\ \bibnamefont {Krasnov}},\
  }\href@noop {} {\bibfield  {journal} {\bibinfo  {journal} {Nat. Commun.}\
  }\textbf {\bibinfo {volume} {6}},\ \bibinfo {pages} {8628} (\bibinfo {year}
  {2015})}\BibitemShut {NoStop}%
\bibitem [{\citenamefont {Najafi}\ \emph {et~al.}(2015)\citenamefont {Najafi},
  \citenamefont {Mower}, \citenamefont {Harris}, \citenamefont {Bellei},
  \citenamefont {Dane}, \citenamefont {Lee}, \citenamefont {Hu}, \citenamefont
  {Kharel}, \citenamefont {Marsili}, \citenamefont {Assefa}, \citenamefont
  {Berggren},\ and\ \citenamefont {Englund}}]{Najafi2015}%
  \BibitemOpen
  \bibfield  {author} {\bibinfo {author} {\bibfnamefont {F.}~\bibnamefont
  {Najafi}}, \bibinfo {author} {\bibfnamefont {J.}~\bibnamefont {Mower}},
  \bibinfo {author} {\bibfnamefont {N.~C.}\ \bibnamefont {Harris}}, \bibinfo
  {author} {\bibfnamefont {F.}~\bibnamefont {Bellei}}, \bibinfo {author}
  {\bibfnamefont {A.}~\bibnamefont {Dane}}, \bibinfo {author} {\bibfnamefont
  {C.}~\bibnamefont {Lee}}, \bibinfo {author} {\bibfnamefont {X.}~\bibnamefont
  {Hu}}, \bibinfo {author} {\bibfnamefont {P.}~\bibnamefont {Kharel}}, \bibinfo
  {author} {\bibfnamefont {F.}~\bibnamefont {Marsili}}, \bibinfo {author}
  {\bibfnamefont {S.}~\bibnamefont {Assefa}}, \bibinfo {author} {\bibfnamefont
  {K.~K.}\ \bibnamefont {Berggren}}, \ and\ \bibinfo {author} {\bibfnamefont
  {D.}~\bibnamefont {Englund}},\ }\href {http://dx.doi.org/10.1038/ncomms6873}
  {\bibfield  {journal} {\bibinfo  {journal} {Nat. Commun.}\ }\textbf {\bibinfo
  {volume} {6}},\ \bibinfo {pages} {5873} (\bibinfo {year} {2015})}\BibitemShut
  {NoStop}%
\bibitem [{\citenamefont {Lowell}\ \emph {et~al.}(2016)\citenamefont {Lowell},
  \citenamefont {Mates}, \citenamefont {Doriese}, \citenamefont {Hilton},
  \citenamefont {Morgan}, \citenamefont {Swetz}, \citenamefont {Ullom},\ and\
  \citenamefont {Schmidt}}]{Lowell2016}%
  \BibitemOpen
  \bibfield  {author} {\bibinfo {author} {\bibfnamefont {P.~J.}\ \bibnamefont
  {Lowell}}, \bibinfo {author} {\bibfnamefont {J.~A.~B.}\ \bibnamefont
  {Mates}}, \bibinfo {author} {\bibfnamefont {W.~B.}\ \bibnamefont {Doriese}},
  \bibinfo {author} {\bibfnamefont {G.~C.}\ \bibnamefont {Hilton}}, \bibinfo
  {author} {\bibfnamefont {K.~M.}\ \bibnamefont {Morgan}}, \bibinfo {author}
  {\bibfnamefont {D.~S.}\ \bibnamefont {Swetz}}, \bibinfo {author}
  {\bibfnamefont {J.~N.}\ \bibnamefont {Ullom}}, \ and\ \bibinfo {author}
  {\bibfnamefont {D.~R.}\ \bibnamefont {Schmidt}},\ }\href {\doibase
  http://dx.doi.org/10.1063/1.4964345} {\bibfield  {journal} {\bibinfo
  {journal} {Appl. Phys. Lett.}\ }\textbf {\bibinfo {volume} {109}},\ \bibinfo
  {pages} {142601} (\bibinfo {year} {2016})}\BibitemShut {NoStop}%
\bibitem [{\citenamefont {Saniz}\ \emph {et~al.}(2013)\citenamefont {Saniz},
  \citenamefont {Partoens},\ and\ \citenamefont
  {Peeters}}]{PhysRevB.87.064510}%
  \BibitemOpen
  \bibfield  {author} {\bibinfo {author} {\bibfnamefont {R.}~\bibnamefont
  {Saniz}}, \bibinfo {author} {\bibfnamefont {B.}~\bibnamefont {Partoens}}, \
  and\ \bibinfo {author} {\bibfnamefont {F.~M.}\ \bibnamefont {Peeters}},\
  }\href {\doibase 10.1103/PhysRevB.87.064510} {\bibfield  {journal} {\bibinfo
  {journal} {Phys. Rev. B}\ }\textbf {\bibinfo {volume} {87}},\ \bibinfo
  {pages} {064510} (\bibinfo {year} {2013})}\BibitemShut {NoStop}%
\bibitem [{\citenamefont {Tang}\ and\ \citenamefont
  {Ismail-Beigi}(2009)}]{PhysRevB.80.134113}%
  \BibitemOpen
  \bibfield  {author} {\bibinfo {author} {\bibfnamefont {H.}~\bibnamefont
  {Tang}}\ and\ \bibinfo {author} {\bibfnamefont {S.}~\bibnamefont
  {Ismail-Beigi}},\ }\href {\doibase 10.1103/PhysRevB.80.134113} {\bibfield
  {journal} {\bibinfo  {journal} {Phys. Rev. B}\ }\textbf {\bibinfo {volume}
  {80}},\ \bibinfo {pages} {134113} (\bibinfo {year} {2009})}\BibitemShut
  {NoStop}%
\bibitem [{\citenamefont {Cepek}\ \emph {et~al.}(2004)\citenamefont {Cepek},
  \citenamefont {Macovez}, \citenamefont {Sancrotti}, \citenamefont {Petaccia},
  \citenamefont {Larciprete}, \citenamefont {Lizzit},\ and\ \citenamefont
  {Goldoni}}]{Cepek2004}%
  \BibitemOpen
  \bibfield  {author} {\bibinfo {author} {\bibfnamefont {C.}~\bibnamefont
  {Cepek}}, \bibinfo {author} {\bibfnamefont {R.}~\bibnamefont {Macovez}},
  \bibinfo {author} {\bibfnamefont {M.}~\bibnamefont {Sancrotti}}, \bibinfo
  {author} {\bibfnamefont {L.}~\bibnamefont {Petaccia}}, \bibinfo {author}
  {\bibfnamefont {R.}~\bibnamefont {Larciprete}}, \bibinfo {author}
  {\bibfnamefont {S.}~\bibnamefont {Lizzit}}, \ and\ \bibinfo {author}
  {\bibfnamefont {A.}~\bibnamefont {Goldoni}},\ }\href@noop {} {\bibfield
  {journal} {\bibinfo  {journal} {Applied Physics Letters}\ }\textbf {\bibinfo
  {volume} {85}},\ \bibinfo {pages} {976} (\bibinfo {year} {2004})}\BibitemShut
  {NoStop}%
\bibitem [{\citenamefont {Sza\l{}owski}(2006)}]{PhysRevB.74.094501}%
  \BibitemOpen
  \bibfield  {author} {\bibinfo {author} {\bibfnamefont {K.}~\bibnamefont
  {Sza\l{}owski}},\ }\href {\doibase 10.1103/PhysRevB.74.094501} {\bibfield
  {journal} {\bibinfo  {journal} {Phys. Rev. B}\ }\textbf {\bibinfo {volume}
  {74}},\ \bibinfo {pages} {094501} (\bibinfo {year} {2006})}\BibitemShut
  {NoStop}%
\bibitem [{\citenamefont {Pogrebnyakov}\ \emph {et~al.}(2004)\citenamefont
  {Pogrebnyakov}, \citenamefont {Redwing}, \citenamefont {Raghavan},
  \citenamefont {Vaithyanathan}, \citenamefont {Schlom}, \citenamefont {Xu},
  \citenamefont {Li}, \citenamefont {Tenne}, \citenamefont {Soukiassian},
  \citenamefont {Xi}, \citenamefont {Johannes}, \citenamefont {Kasinathan},
  \citenamefont {Pickett}, \citenamefont {Wu},\ and\ \citenamefont
  {Spence}}]{PhysRevLett.93.147006}%
  \BibitemOpen
  \bibfield  {author} {\bibinfo {author} {\bibfnamefont {A.~V.}\ \bibnamefont
  {Pogrebnyakov}}, \bibinfo {author} {\bibfnamefont {J.~M.}\ \bibnamefont
  {Redwing}}, \bibinfo {author} {\bibfnamefont {S.}~\bibnamefont {Raghavan}},
  \bibinfo {author} {\bibfnamefont {V.}~\bibnamefont {Vaithyanathan}}, \bibinfo
  {author} {\bibfnamefont {D.~G.}\ \bibnamefont {Schlom}}, \bibinfo {author}
  {\bibfnamefont {S.~Y.}\ \bibnamefont {Xu}}, \bibinfo {author} {\bibfnamefont
  {Q.}~\bibnamefont {Li}}, \bibinfo {author} {\bibfnamefont {D.~A.}\
  \bibnamefont {Tenne}}, \bibinfo {author} {\bibfnamefont {A.}~\bibnamefont
  {Soukiassian}}, \bibinfo {author} {\bibfnamefont {X.~X.}\ \bibnamefont {Xi}},
  \bibinfo {author} {\bibfnamefont {M.~D.}\ \bibnamefont {Johannes}}, \bibinfo
  {author} {\bibfnamefont {D.}~\bibnamefont {Kasinathan}}, \bibinfo {author}
  {\bibfnamefont {W.~E.}\ \bibnamefont {Pickett}}, \bibinfo {author}
  {\bibfnamefont {J.~S.}\ \bibnamefont {Wu}}, \ and\ \bibinfo {author}
  {\bibfnamefont {J.~C.~H.}\ \bibnamefont {Spence}},\ }\href {\doibase
  10.1103/PhysRevLett.93.147006} {\bibfield  {journal} {\bibinfo  {journal}
  {Phys. Rev. Lett.}\ }\textbf {\bibinfo {volume} {93}},\ \bibinfo {pages}
  {147006} (\bibinfo {year} {2004})}\BibitemShut {NoStop}%
\bibitem [{\citenamefont {Zheng}\ and\ \citenamefont
  {Zhu}(2006)}]{PhysRevB.73.024509}%
  \BibitemOpen
  \bibfield  {author} {\bibinfo {author} {\bibfnamefont {J.-C.}\ \bibnamefont
  {Zheng}}\ and\ \bibinfo {author} {\bibfnamefont {Y.}~\bibnamefont {Zhu}},\
  }\href {\doibase 10.1103/PhysRevB.73.024509} {\bibfield  {journal} {\bibinfo
  {journal} {Phys. Rev. B}\ }\textbf {\bibinfo {volume} {73}},\ \bibinfo
  {pages} {024509} (\bibinfo {year} {2006})}\BibitemShut {NoStop}%
\bibitem [{\citenamefont {Pe\v{s}i\'{c}}\ \emph {et~al.}(2014)\citenamefont
  {Pe\v{s}i\'{c}}, \citenamefont {Gaji\'{c}}, \citenamefont {Hingerl},\ and\
  \citenamefont {Beli\'{c}}}]{0295-5075-108-6-67005}%
  \BibitemOpen
  \bibfield  {author} {\bibinfo {author} {\bibfnamefont {J.}~\bibnamefont
  {Pe\v{s}i\'{c}}}, \bibinfo {author} {\bibfnamefont {R.}~\bibnamefont
  {Gaji\'{c}}}, \bibinfo {author} {\bibfnamefont {K.}~\bibnamefont {Hingerl}},
  \ and\ \bibinfo {author} {\bibfnamefont {M.}~\bibnamefont {Beli\'{c}}},\
  }\href {http://stacks.iop.org/0295-5075/108/i=6/a=67005} {\bibfield
  {journal} {\bibinfo  {journal} {EPL}\ }\textbf {\bibinfo {volume} {108}},\
  \bibinfo {pages} {67005} (\bibinfo {year} {2014})}\BibitemShut {NoStop}%
\bibitem [{\citenamefont {Si}\ \emph {et~al.}(2013)\citenamefont {Si},
  \citenamefont {Liu}, \citenamefont {Duan},\ and\ \citenamefont
  {Liu}}]{PhysRevLett.111.196802}%
  \BibitemOpen
  \bibfield  {author} {\bibinfo {author} {\bibfnamefont {C.}~\bibnamefont
  {Si}}, \bibinfo {author} {\bibfnamefont {Z.}~\bibnamefont {Liu}}, \bibinfo
  {author} {\bibfnamefont {W.}~\bibnamefont {Duan}}, \ and\ \bibinfo {author}
  {\bibfnamefont {F.}~\bibnamefont {Liu}},\ }\href {\doibase
  10.1103/PhysRevLett.111.196802} {\bibfield  {journal} {\bibinfo  {journal}
  {Phys. Rev. Lett.}\ }\textbf {\bibinfo {volume} {111}},\ \bibinfo {pages}
  {196802} (\bibinfo {year} {2013})}\BibitemShut {NoStop}%
\bibitem [{\citenamefont {Si}\ \emph {et~al.}(2016)\citenamefont {Si},
  \citenamefont {Sun},\ and\ \citenamefont {Liu}}]{C5NR07755A}%
  \BibitemOpen
  \bibfield  {author} {\bibinfo {author} {\bibfnamefont {C.}~\bibnamefont
  {Si}}, \bibinfo {author} {\bibfnamefont {Z.}~\bibnamefont {Sun}}, \ and\
  \bibinfo {author} {\bibfnamefont {F.}~\bibnamefont {Liu}},\ }\href {\doibase
  10.1039/C5NR07755A} {\bibfield  {journal} {\bibinfo  {journal} {Nanoscale}\
  }\textbf {\bibinfo {volume} {8}},\ \bibinfo {pages} {3207} (\bibinfo {year}
  {2016})}\BibitemShut {NoStop}%
\bibitem [{\citenamefont {Gonze}\ \emph {et~al.}(2009)\citenamefont {Gonze},
  \citenamefont {Amadon}, \citenamefont {Anglade}, \citenamefont {Beuken},
  \citenamefont {Bottin}, \citenamefont {Boulanger}, \citenamefont {Bruneval},
  \citenamefont {Caliste}, \citenamefont {Caracas}, \citenamefont
  {C\^{o}t\'{e}}, \citenamefont {Deutsch}, \citenamefont {Genovese},
  \citenamefont {Ghosez}, \citenamefont {Giantomassi}, \citenamefont
  {Goedecker}, \citenamefont {Hamann}, \citenamefont {Hermet}, \citenamefont
  {Jollet}, \citenamefont {Jomard}, \citenamefont {Leroux}, \citenamefont
  {Mancini}, \citenamefont {Mazevet}, \citenamefont {Oliveira}, \citenamefont
  {Onida}, \citenamefont {Pouillon}, \citenamefont {Rangel}, \citenamefont
  {Rignanese}, \citenamefont {Sangalli}, \citenamefont {Shaltaf}, \citenamefont
  {Torrent}, \citenamefont {Verstraete}, \citenamefont {Zerah},\ and\
  \citenamefont {Zwanziger}}]{Gonze20092582}%
  \BibitemOpen
  \bibfield  {author} {\bibinfo {author} {\bibfnamefont {X.}~\bibnamefont
  {Gonze}}, \bibinfo {author} {\bibfnamefont {B.}~\bibnamefont {Amadon}},
  \bibinfo {author} {\bibfnamefont {P.-M.}\ \bibnamefont {Anglade}}, \bibinfo
  {author} {\bibfnamefont {J.-M.}\ \bibnamefont {Beuken}}, \bibinfo {author}
  {\bibfnamefont {F.}~\bibnamefont {Bottin}}, \bibinfo {author} {\bibfnamefont
  {P.}~\bibnamefont {Boulanger}}, \bibinfo {author} {\bibfnamefont
  {F.}~\bibnamefont {Bruneval}}, \bibinfo {author} {\bibfnamefont
  {D.}~\bibnamefont {Caliste}}, \bibinfo {author} {\bibfnamefont
  {R.}~\bibnamefont {Caracas}}, \bibinfo {author} {\bibfnamefont
  {M.}~\bibnamefont {C\^{o}t\'{e}}}, \bibinfo {author} {\bibfnamefont
  {T.}~\bibnamefont {Deutsch}}, \bibinfo {author} {\bibfnamefont
  {L.}~\bibnamefont {Genovese}}, \bibinfo {author} {\bibfnamefont
  {P.}~\bibnamefont {Ghosez}}, \bibinfo {author} {\bibfnamefont
  {M.}~\bibnamefont {Giantomassi}}, \bibinfo {author} {\bibfnamefont
  {S.}~\bibnamefont {Goedecker}}, \bibinfo {author} {\bibfnamefont
  {D.}~\bibnamefont {Hamann}}, \bibinfo {author} {\bibfnamefont
  {P.}~\bibnamefont {Hermet}}, \bibinfo {author} {\bibfnamefont
  {F.}~\bibnamefont {Jollet}}, \bibinfo {author} {\bibfnamefont
  {G.}~\bibnamefont {Jomard}}, \bibinfo {author} {\bibfnamefont
  {S.}~\bibnamefont {Leroux}}, \bibinfo {author} {\bibfnamefont
  {M.}~\bibnamefont {Mancini}}, \bibinfo {author} {\bibfnamefont
  {S.}~\bibnamefont {Mazevet}}, \bibinfo {author} {\bibfnamefont
  {M.}~\bibnamefont {Oliveira}}, \bibinfo {author} {\bibfnamefont
  {G.}~\bibnamefont {Onida}}, \bibinfo {author} {\bibfnamefont
  {Y.}~\bibnamefont {Pouillon}}, \bibinfo {author} {\bibfnamefont
  {T.}~\bibnamefont {Rangel}}, \bibinfo {author} {\bibfnamefont {G.-M.}\
  \bibnamefont {Rignanese}}, \bibinfo {author} {\bibfnamefont {D.}~\bibnamefont
  {Sangalli}}, \bibinfo {author} {\bibfnamefont {R.}~\bibnamefont {Shaltaf}},
  \bibinfo {author} {\bibfnamefont {M.}~\bibnamefont {Torrent}}, \bibinfo
  {author} {\bibfnamefont {M.}~\bibnamefont {Verstraete}}, \bibinfo {author}
  {\bibfnamefont {G.}~\bibnamefont {Zerah}}, \ and\ \bibinfo {author}
  {\bibfnamefont {J.}~\bibnamefont {Zwanziger}},\ }\href {\doibase
  http://dx.doi.org/10.1016/j.cpc.2009.07.007} {\bibfield  {journal} {\bibinfo
  {journal} {Comput. Phys. Commun.}\ }\textbf {\bibinfo {volume} {180}},\
  \bibinfo {pages} {2582 } (\bibinfo {year} {2009})}\BibitemShut {NoStop}%
\bibitem [{sup()}]{suppl_mat}%
  \BibitemOpen
  \href@noop {} {\bibinfo  {journal} {See Supplemental Material at [URL will be
  inserted by publisher]}\ }\BibitemShut {NoStop}%
\bibitem [{\citenamefont {Savrasov}\ and\ \citenamefont
  {Savrasov}(1996)}]{PhysRevB.54.16487}%
  \BibitemOpen
\bibfield  {journal} {  }\bibfield  {author} {\bibinfo {author} {\bibfnamefont
  {S.~Y.}\ \bibnamefont {Savrasov}}\ and\ \bibinfo {author} {\bibfnamefont
  {D.~Y.}\ \bibnamefont {Savrasov}},\ }\href {\doibase
  10.1103/PhysRevB.54.16487} {\bibfield  {journal} {\bibinfo  {journal} {Phys.
  Rev. B}\ }\textbf {\bibinfo {volume} {54}},\ \bibinfo {pages} {16487}
  (\bibinfo {year} {1996})}\BibitemShut {NoStop}%
\bibitem [{\citenamefont {Bekaert}\ \emph {et~al.}(2016)\citenamefont
  {Bekaert}, \citenamefont {Vercauteren}, \citenamefont {Aperis}, \citenamefont
  {Komendov\'a}, \citenamefont {Prozorov}, \citenamefont {Partoens},\ and\
  \citenamefont {Milo\ifmmode \check{s}\else
  \v{s}\fi{}evi\ifmmode~\acute{c}\else \'{c}\fi{}}}]{PhysRevB.94.144506}%
  \BibitemOpen
  \bibfield  {author} {\bibinfo {author} {\bibfnamefont {J.}~\bibnamefont
  {Bekaert}}, \bibinfo {author} {\bibfnamefont {S.}~\bibnamefont
  {Vercauteren}}, \bibinfo {author} {\bibfnamefont {A.}~\bibnamefont {Aperis}},
  \bibinfo {author} {\bibfnamefont {L.}~\bibnamefont {Komendov\'a}}, \bibinfo
  {author} {\bibfnamefont {R.}~\bibnamefont {Prozorov}}, \bibinfo {author}
  {\bibfnamefont {B.}~\bibnamefont {Partoens}}, \ and\ \bibinfo {author}
  {\bibfnamefont {M.~V.}\ \bibnamefont {Milo\ifmmode \check{s}\else
  \v{s}\fi{}evi\ifmmode~\acute{c}\else \'{c}\fi{}}},\ }\href {\doibase
  10.1103/PhysRevB.94.144506} {\bibfield  {journal} {\bibinfo  {journal} {Phys.
  Rev. B}\ }\textbf {\bibinfo {volume} {94}},\ \bibinfo {pages} {144506}
  (\bibinfo {year} {2016})}\BibitemShut {NoStop}%
\bibitem [{\citenamefont {Moon}\ \emph {et~al.}(2004)\citenamefont {Moon},
  \citenamefont {Kim},\ and\ \citenamefont {Chang}}]{PhysRevB.70.104522}%
  \BibitemOpen
  \bibfield  {author} {\bibinfo {author} {\bibfnamefont {C.-Y.}\ \bibnamefont
  {Moon}}, \bibinfo {author} {\bibfnamefont {Y.-H.}\ \bibnamefont {Kim}}, \
  and\ \bibinfo {author} {\bibfnamefont {K.~J.}\ \bibnamefont {Chang}},\ }\href
  {\doibase 10.1103/PhysRevB.70.104522} {\bibfield  {journal} {\bibinfo
  {journal} {Phys. Rev. B}\ }\textbf {\bibinfo {volume} {70}},\ \bibinfo
  {pages} {104522} (\bibinfo {year} {2004})}\BibitemShut {NoStop}%
\bibitem [{Note1()}]{Note1}%
  \BibitemOpen
  \bibinfo {note} {We note here that this result is different from that
  obtained in Ref.~\protect \citenum {MORSHEDLOO20151} for 2-ML MgB$_2$, where
  $T_{\protect \mathrm {c}}$ was found to exceed the bulk value. The difference
  can be traced back to the unreasonably low Coulomb pseudopotential used in
  this work, to compensate the lack of multi-band effects in their isotropic
  Eliashberg approach.}\BibitemShut {Stop}%
\bibitem [{\citenamefont {Souma}\ \emph {et~al.}(2003)\citenamefont {Souma},
  \citenamefont {Machida}, \citenamefont {Sato}, \citenamefont {Takahashi},
  \citenamefont {Wang}, \citenamefont {Ding}, \citenamefont {Kaminski},
  \citenamefont {Campuzano}, \citenamefont {Sasaki},\ and\ \citenamefont
  {Kadowaki}}]{Souma03}%
  \BibitemOpen
  \bibfield  {author} {\bibinfo {author} {\bibfnamefont {S.}~\bibnamefont
  {Souma}}, \bibinfo {author} {\bibfnamefont {Y.}~\bibnamefont {Machida}},
  \bibinfo {author} {\bibfnamefont {T.}~\bibnamefont {Sato}}, \bibinfo {author}
  {\bibfnamefont {T.}~\bibnamefont {Takahashi}}, \bibinfo {author}
  {\bibfnamefont {H.~M. S.-C.}\ \bibnamefont {Wang}}, \bibinfo {author}
  {\bibfnamefont {H.}~\bibnamefont {Ding}}, \bibinfo {author} {\bibfnamefont
  {A.}~\bibnamefont {Kaminski}}, \bibinfo {author} {\bibfnamefont {J.~C.}\
  \bibnamefont {Campuzano}}, \bibinfo {author} {\bibfnamefont {S.}~\bibnamefont
  {Sasaki}}, \ and\ \bibinfo {author} {\bibfnamefont {K.}~\bibnamefont
  {Kadowaki}},\ }\href@noop {} {\bibfield  {journal} {\bibinfo  {journal}
  {Nature}\ }\textbf {\bibinfo {volume} {423}},\ \bibinfo {pages} {65}
  (\bibinfo {year} {2003})}\BibitemShut {NoStop}%
\bibitem [{\citenamefont {Grimvall}(1981)}]{Grimvall}%
  \BibitemOpen
  \bibfield  {author} {\bibinfo {author} {\bibfnamefont {G.}~\bibnamefont
  {Grimvall}},\ }\href@noop {} {\emph {\bibinfo {title} {The electron-phonon
  interaction}}}\ (\bibinfo  {publisher} {North Holland Publishing Co.},\
  \bibinfo {year} {1981})\BibitemShut {NoStop}%
\bibitem [{Note2()}]{Note2}%
  \BibitemOpen
  \bibinfo {note} {We note that for compressive strains exceeding $-1.5$\%
  $\sigma $- and \protect \textit {S}-gaps become hybridized, albeit their
  contributions can still be distinguished. Their partial overlap is not due to
  new physics -- it is provoked by a general depletion of the superconducting
  gap values, forcing the gaps closer together.}\BibitemShut {Stop}%
\bibitem [{\citenamefont {Hamann}(2013)}]{PhysRevB.88.085117}%
  \BibitemOpen
  \bibfield  {author} {\bibinfo {author} {\bibfnamefont {D.~R.}\ \bibnamefont
  {Hamann}},\ }\href {\doibase 10.1103/PhysRevB.88.085117} {\bibfield
  {journal} {\bibinfo  {journal} {Phys. Rev. B}\ }\textbf {\bibinfo {volume}
  {88}},\ \bibinfo {pages} {085117} (\bibinfo {year} {2013})}\BibitemShut
  {NoStop}%
\bibitem [{\citenamefont {Beach}\ \emph {et~al.}(2000)\citenamefont {Beach},
  \citenamefont {Gooding},\ and\ \citenamefont {Marsiglio}}]{PhysRevB.61.5147}%
  \BibitemOpen
  \bibfield  {author} {\bibinfo {author} {\bibfnamefont {K.~S.~D.}\
  \bibnamefont {Beach}}, \bibinfo {author} {\bibfnamefont {R.~J.}\ \bibnamefont
  {Gooding}}, \ and\ \bibinfo {author} {\bibfnamefont {F.}~\bibnamefont
  {Marsiglio}},\ }\href {\doibase 10.1103/PhysRevB.61.5147} {\bibfield
  {journal} {\bibinfo  {journal} {Phys. Rev. B}\ }\textbf {\bibinfo {volume}
  {61}},\ \bibinfo {pages} {5147} (\bibinfo {year} {2000})}\BibitemShut
  {NoStop}%
\bibitem [{\citenamefont {Morshedloo}\ \emph {et~al.}(2015)\citenamefont
  {Morshedloo}, \citenamefont {Roknabadi},\ and\ \citenamefont
  {Behdani}}]{MORSHEDLOO20151}%
  \BibitemOpen
  \bibfield  {author} {\bibinfo {author} {\bibfnamefont {T.}~\bibnamefont
  {Morshedloo}}, \bibinfo {author} {\bibfnamefont {M.}~\bibnamefont
  {Roknabadi}}, \ and\ \bibinfo {author} {\bibfnamefont {M.}~\bibnamefont
  {Behdani}},\ }\href {\doibase http://dx.doi.org/10.1016/j.physc.2014.11.006}
  {\bibfield  {journal} {\bibinfo  {journal} {Physica C Supercond.}\ }\textbf
  {\bibinfo {volume} {509}},\ \bibinfo {pages} {1 } (\bibinfo {year}
  {2015})}\BibitemShut {NoStop}%
\end{thebibliography}%

\end{document}